\documentclass[aps,prd,twocolumn,superscriptaddress,nofootinbib,bibliography]{revtex4-2}

\usepackage[T1]{fontenc}
\usepackage[utf8]{inputenc}
\usepackage{lmodern}
\usepackage{amsmath,amssymb,mathtools,bm}
\usepackage{microtype}
\usepackage{tikz}
\usetikzlibrary{calc,arrows.meta,positioning}
\usepackage{comment}
\usepackage[colorlinks=true,citecolor=blue,linkcolor=blue,urlcolor=blue]{hyperref}
\allowdisplaybreaks

\newcommand{\dd}{\mathrm d}
\newcommand{\ee}{\mathrm e}

\newcommand{\cP}{\mathcal P}
\newcommand{\hP}{\widehat{\mathcal P}}
\newcommand{\hPF}{\widehat{\mathcal P}_{F}}
\newcommand{\hPf}{\widehat{\mathcal P}_{f}}
\newcommand{\cN}{\mathcal N}
\newcommand{\cM}{\mathcal M}
\newcommand{\cL}{\mathcal L}
\newcommand{\cF}{\mathcal F}

\newcommand{\cO}{\mathcal O}
\newcommand{\BoxE}{\widetilde\Box}
\newcommand{\nablaE}{\widetilde\nabla}
\newcommand{\gE}{\widetilde g}
\newcommand{\RE}{\widetilde R}
\newcommand{\kapp}{\kappa}

\newcommand{\Lie}{\mathcal L}

\newcommand{\Tr}{\operatorname{Tr}}

\definecolor{lime}{HTML}{A6CE39}
\DeclareRobustCommand{\orcidicon}{%
 \begin{tikzpicture}
 \draw[lime, fill=lime] (0,0) 
 circle [radius=0.16] 
 node[white] {{\fontfamily{qag}\selectfont \tiny ID}};
 \draw[white, fill=white] (-0.0625,0.095) 
 circle [radius=0.007];
 \end{tikzpicture}
 \hspace{-2mm}
}

\foreach \x in {A, ..., Z}{%
 \expandafter\xdef\csname orcid\x\endcsname{\noexpand\href{https://orcid.org/\csname orcidauthor\x\endcsname}{\noexpand\orcidicon}}
}

\newcommand\orcidDavid{{\href{https://orcid.org/0009-0006-8485-3140}{\orcidicon}}}
\newcommand\orcidFrancisco{{\href{https://orcid.org/0000-0002-9388-8373}{\orcidicon}}}
\newcommand\orcidJPedro{{\href{https://orcid.org/0000-0002-9758-3366}{\orcidicon}}}

\begin{document}

\title{Differential Obstructions to Curvature-Dependent Conformal Transformations}

\author{David S. Pereira\orcidDavid\!\!}
\email{djpereira@ciencias.ulisboa.pt}

\author{Francisco S.N. Lobo\orcidFrancisco\!\!}
\email{fslobo@ciencias.ulisboa.pt}

\author{Jos\'e Pedro Mimoso\orcidJPedro\!\!}
\email{jpmimoso@ciencias.ulisboa.pt}

\affiliation{Departamento de F\'{i}sica, Faculdade de Ci\^{e}ncias da Universidade de Lisboa, Campo Grande, Edif\'{i}cio C8, P-1749-016 Lisbon, Portugal}
\affiliation{Instituto de Astrof\'{i}sica e Ci\^{e}ncias do Espa\c{c}o, Faculdade de Ci\^{e}ncias da Universidade de Lisboa, Campo Grande, P-1749-016 Lisbon, Portugal}

\date{\today}

\begin{abstract}
Curvature-dependent conformal rules are local forward assignments on known metrics, but they are not generically local changes of metric variables. We study the nondegenerate class $\widetilde g_{\mu\nu}=F(R[g])g_{\mu\nu}$, with $F>0$ and $F_R\neq0$. Introducing an independent auxiliary scalar localizes the forward map, while recovering the original metric requires a differential constraint. The complete inverse metric tangent map contains the nonpolynomial projector $\xi^\mu\xi^\nu/\xi^2$; hence no differentiable finite-jet inverse, i.e., a formula involving only finitely many derivatives at the same point, exists on an open set of unrestricted metrics. Branchwise functional inverses may nevertheless exist after boundary or Cauchy data are specified. Metric $f(R)$ gravity gives an explicit realization: its local Einstein-frame scalar--tensor representation is a parent theory, whereas a metric-only Einstein-side description requires a differential section governed by a normal operator. We derive the pulled-back classical Hessian and its off-shell embedding term, and in the quadratic model show how constrained Gaussian elimination produces the scalaron nonlocal kernel, the corresponding normal determinant for the displayed measure, and the zero-mode compatibility condition. Exact parent and metric solutions remain equivalent; the obstruction concerns locality and the off-shell variational and fluctuation domains. These results provide a precise framework for assessing curvature-dependent frame transformations in modified gravity and clarify their implications for effective actions, semiclassical analyses, and quantum frame equivalence.
\end{abstract}

\maketitle


\section{Introduction}

Conformal transformations are among the most useful tools in gravitational theory. In scalar-tensor models one often introduces an Einstein-frame metric through an appropriate rescaling of the metric
\begin{equation}
	\gE_{\mu\nu}=\Omega^2(\phi)g_{\mu\nu},
	\label{eq:intro_scalar_conformal}
\end{equation}
where the conformal factor $\Omega(\phi)$ is a function of an independent scalar field, and is prescribed to recover the Einstein-Hilbert geometric sector of the fundamental action. 

Wherever $\Omega(\phi)$ is finite and nonzero, the inverse transformation to \eqref{eq:intro_scalar_conformal} is algebraic: $g_{\mu\nu}=\Omega^{-2}(\phi)\gE_{\mu\nu}$.
The two sets of variables therefore provide local descriptions of the same enlarged field space. This is the familiar setting in which conformal-frame comparisons and field-redefinition equivalence theorems are usually discussed \cite{Chisholm:1961tha,Kamefuchi:1961sb,Dicke:1961gz,Faraoni:1998qx, FujiiMaedaBook,Flanagan:2004bz,Faraoni:2006fx,Cotsakis:2023uyt}.
 
The situation is more subtle when the conformal factor is not an independent field, but is instead constructed from the curvature of the metric being transformed. The class considered in this work is
 \begin{equation}
 \gE_{\mu\nu}=F(R[g])g_{\mu\nu},
 \qquad F(R)>0,
 \label{eq:intro_curvature_conformal}
 \end{equation}
on a smooth, nondegenerate branch with $F(R)$ being a generic function of the Ricci scalar that satisfies $F_R\neq0$. Such transformations arise naturally in modified gravity, reconstruction procedures, solution-generating methods, and attempts to formulate higher-curvature theories in variables resembling those of general relativity.

For a known metric $g_{\mu\nu}$, the forward transformation is straightforward: one calculates $R[g]$, evaluates $F(R[g])$, and rescales the metric. The inverse problem is fundamentally different. If $\gE_{\mu\nu}$ is given, the conformal factor still depends on the curvature of the unknown original metric. One must therefore determine the metric and its curvature self-consistently. The inverse is not obtained by the simple replacement $R[g]\rightarrow\RE$.
 
This observation raises the main question addressed in this paper:
 \begin{quote}
 Can a curvature-dependent conformal transformation be regarded as a genuine local change of metric variables, in the same sense as an ordinary
 scalar--tensor conformal transformation?
 \end{quote}
By ``local'' we mean that the original metric can be reconstructed at each spacetime point from the transformed metric and finitely many of its derivatives at that same point. This is often called \emph{finite-jet locality}. A transformation that instead requires solving a differential equation with prescribed boundary or initial data is not local in this sense. Hence by non-local we mean that the inverse cannot be expressed as a finite-jet functional of the transformed metric, but instead requires the action of an inverse differential operator, or equivalently a Green operator, whose definition depends on the chosen functional domain and boundary or Cauchy prescription. Nonlocality in this sense does not imply acausality: for example, in Lorentzian signature the inverse may be defined by a retarded Green operator.

The distinction is important because the forward relation in Eq.~\eqref{eq:intro_curvature_conformal} can look deceptively similar to the ordinary scalar--tensor transformation in Eq.~\eqref{eq:intro_scalar_conformal}. Nevertheless, the two constructions act on different field spaces. In the scalar--tensor case, the conformal factor is an independent coordinate. In the curvature-dependent case, it is already a functional of the metric and therefore carries hidden differential information.
 
Derivative dependence alone does not prove that a transformation lacks a local inverse. Special derivative-dependent transformations may still be invertible through finite-order relations \cite{Babichev:2019twf, Babichev:2021bim}. A definitive conclusion must therefore follow from the complete inverse response, rather than merely from the appearance of derivatives in the forward map.
 
We show that, for the nondegenerate class in Eq.~\eqref{eq:intro_curvature_conformal}, the complete inverse metric response contains an inverse differential operator. Equivalently, its local momentum-space representation contains the nonpolynomial factor $1/\xi^2$. This factor signals that reconstructing the original metric requires solving a differential problem rather than applying finitely many local derivatives. Consequently, the transformation does not possess a generic differentiable finite-jet metric inverse on an open set of unrestricted metric configurations.
 
This conclusion does not mean that the transformation can never be inverted. Branchwise functional inverses may exist after a functional domain, boundary or Cauchy data, and an appropriate solution prescription have been specified. The essential result is therefore a statement about the \emph{locality} and \emph{off-shell status} of the inverse, rather than an assertion that every form of inversion is impossible.
 
Metric $f(R)$ gravity provides a particularly important realization of this distinction. On a regular Legendre branch, the theory can be written as a local scalar--tensor theory and then transformed to the Einstein frame \cite{Whitt:1984pd,Maeda:1988ab,Teyssandier:1983zz,Wands:1993uu,Magnano:1993bd,Sotiriou:2008rp,DeFelice:2010aj}. In that enlarged formulation, the scalar field is independent and the conformal transformation is local. The original metric theory, however, is recovered only after the scalar is identified with the curvature carried by the higher-derivative metric sector. In Einstein-frame variables, this identification becomes a differential constraint.
 
The familiar scalar--tensor Einstein-frame action should therefore be understood as a local \emph{parent theory}. A formulation involving the Einstein-side metric alone is obtained only after solving the differential constraint that selects the metric $f(R)$ configurations inside the parent field space. This difference is invisible when one transforms a known exact solution, but it becomes essential when the transformed variables are treated as an independent off-shell variational or fluctuation space.
 
The result does not alter the standard classical equivalence between regular solutions of metric $f(R)$ gravity and those of its scalar--tensor representation. Exact solutions of the parent theory satisfy the required constraint. The distinction arises away from the common solution space, where arbitrary parent configurations need not correspond to configurations of the original metric theory.
 
This off-shell distinction is relevant to effective actions and semiclassical comparisons. Existing one-loop analyses find that the metric and scalar--tensor formulations may differ off shell, while agreement is recovered after imposing the common equations of motion \cite{Kamenshchik:2014waa,RufSteinwachs2018a,Ruf:2017xon,Ohta:2017trn,Falls:2018olk}. Recent work has further emphasized that the auxiliary-field formulation reproduces the metric quantum theory only when the auxiliary constraint is retained in the functional integral \cite{Kuntz:2026vhs}. The present analysis clarifies the differential structure underlying that constraint and the origin of the associated nonlocal metric-only response.

Our result therefore implies that curvature-dependent conformal transformations should not generically be treated as ordinary local changes of metric variables. Their use as forward maps of known configurations remains valid, as does a local conformal transformation performed on an appropriately enlarged parent field space. What fails generically is the interpretation of the curvature-dependent rule as a local, unconstrained reparametrization of the original off-shell metric configuration space. Consequently, inverse, variational, perturbative, semiclassical, and quantum constructions must retain the differential projection and the functional data required to solve it. Curvature-dependent conformal transformations used in modified gravity must therefore be assessed according to whether they are being used as forward maps, as transformations of a parent theory, or as genuine changes of off-shell variables.
 
The paper is organized as follows. Section~\ref{sec:curvature_maps} analyzes the inverse problem for the general curvature-dependent map and establishes the finite-jet obstruction. Section~\ref{sec:fr_parent} develops the corresponding parent construction in metric $f(R)$ gravity. Section~\ref{sec:normal_operator} derives the differential operator governing the inverse response. Sections~\ref{sec:boundary_terms} and \ref{sec:hessian} discuss the functional data required to define the inverse and the consequences for the quadratic variational problem. Weak-field and cosmological illustrations are presented in Secs.~\ref{sec:weakfield} and \ref{sec:cosmology}. The remaining sections examine possible failures of the metric-only description, its nonlocal form, and extensions beyond metric $f(R)$ gravity.
 
Throughout the paper, we use the signature $(-+++)$, and the conventions $\Box=g^{\mu\nu}\nabla_\mu\nabla_\nu$, $\widetilde\Box =\widetilde g^{\mu\nu} \widetilde\nabla_\mu\widetilde\nabla_\nu$. Tilded quantities are constructed from $\widetilde g_{\mu\nu}$, whereas untilded curvature quantities refer to the original Jordan-frame metric unless stated otherwise. We use $\kapp^2=8\pi G$, where $G$ is Newton's gravitational constant.

\section{Curvature-dependent conformal maps}
\label{sec:curvature_maps}

\subsection{Forward map and inverse fixed point}

Let $g_{\mu\nu}$ be a Lorentzian metric on a four-dimensional spacetime and consider
\begin{equation}
 \cF_F:g_{\mu\nu}\mapsto
 \gE_{\mu\nu}=F(R[g])g_{\mu\nu},
 \label{eq:generic_curv_map}
\end{equation}
with $F>0$. The forward map depends on the metric and its first two derivatives. Given $\gE_{\mu\nu}$, however, write
\begin{equation}
 g_{\mu\nu}=F(R_J)^{-1}\gE_{\mu\nu},
 \qquad R_J\equiv R[g].
 \label{eq:preimage_metric_generic}
\end{equation}
Self-consistency requires
\begin{equation}
 R_J=R\!\left[F(R_J)^{-1}\gE\right],
 \label{eq:self_consistency_generic}
\end{equation}
which is a differential fixed-point equation, not an algebraic replacement. Here ``fixed point'' means that the unknown curvature appears both as the quantity to be determined and inside the differential operator that determines it. Thus the equation must be solved self-consistently rather than by a pointwise algebraic substitution.

The conformal curvature formula gives
\begin{align}
 R_J={}&F\RE+3F_R\BoxE R_J
 \nonumber\\
 &+3\left(F_{RR}-\frac32\frac{F_R^2}{F}\right)
 (\nablaE R_J)^2,
 \label{eq:expanded_fixed_point_generic}
\end{align}
where all functions are evaluated at $R_J$. Equivalently,
\begin{align}
 T_F[R]={}&F(R)\RE+3F_R(R)\BoxE R
 \nonumber\\
 &+3\left(F_{RR}(R)-\frac32\frac{F_R(R)^2}{F(R)}\right)
 (\nablaE R)^2,
 \label{eq:TF_operator}
\end{align}
and $R_J=T_F[R_J]$. Formal iteration,
\begin{equation}
 R_J=T_F[R_J]=T_F[T_F[R_J]]=\cdots,
 \label{eq:fixed_point_iteration}
\end{equation}
generically generates derivatives of arbitrarily high order. A branch, functional domain, and boundary or Cauchy data are therefore part of the inverse construction.

\subsection{Finite-jet obstruction}

Linearize
\begin{equation}
 \cP_F[\gE,R]=R-T_F[R]
 \label{eq:generic_projection_F}
\end{equation}
about a projected background. Here a ``projected background'' means a pair
 $(\gE_{\mu\nu},R)$ satisfying
\begin{equation}
 \cP_F[\gE,R]=0,
\end{equation}
or equivalently the fixed-point condition $R=T_F[R]$ in
Eq.~\eqref{eq:TF_operator}. This geometric condition need not coincide with any dynamical field equation; throughout this section, ``projected'' refers only to satisfaction of the defining inverse constraint. At fixed $\gE$,
\begin{equation}
 \delta_R\cP_F=
 \left(1-3F_R\BoxE+\cdots\right)\delta R,
 \label{eq:generic_normal_principal}
\end{equation}
where the omitted terms contain fewer derivatives. This already indicates a differential inverse, but it is not by itself a proof of nonlocality: the metric source might contain a compensating factor. The decisive object is the complete response to a metric perturbation.

A weak-curvature example illustrates the distinction. For
 $F=1+2\alpha R$,
\begin{equation}
 (1-6\alpha\BoxE)R_J\simeq\RE,
 \qquad
 R_J\simeq(1-6\alpha\BoxE)^{-1}\RE,
 \label{eq:weak_fixed_point_intro}
\end{equation}
and, when $|6\alpha\BoxE|\ll1$,
\begin{equation}
 R_J\simeq\sum_{n=0}^{\infty}(6\alpha\BoxE)^n\RE.
 \label{eq:weak_derivative_series}
\end{equation}
This is a controlled infinite-derivative expansion, not a finite-jet inverse. For broader discussions of nonlocal and infinite-derivative gravity models, we refer the reader to Refs.~\cite{Koivisto:2008dh,Deser:2013uya,delaCruzDombriz:2008cp, Deffayet:2009ca,Nojiri:2010pw,Shapiro:2008sf,Deser:2007jk, Koivisto:2008xf,Bamba:2008ut,Barnaby:2007ve,Giddings:2007pj,Wetterich:1997bz}.

For comparison, if the conformal factor is an independent scalar,
\begin{equation}
 \gE_{\mu\nu}=\Phi g_{\mu\nu},
 \label{eq:independent_phi_map}
\end{equation}
then
\begin{equation}
 g_{\mu\nu}=\Phi^{-1}\gE_{\mu\nu}
 \label{eq:independent_phi_inverse}
\end{equation}
is algebraic, and
\begin{equation}
 R[g]=\Phi\left[\RE+3\BoxE\ln\Phi
 -\frac32(\nablaE\ln\Phi)^2\right]
 \label{eq:R_independent_phi}
\end{equation}
is finite-jet in the independent variables. The obstruction arises only after the carrier is identified with the curvature of the preimage metric.

\subsection{Auxiliary localization and the inverse symbol}
\label{subsec:generic_auxiliary_normal_operator}
By ``auxiliary localization'' we mean temporarily treating the quantity that carries the conformal factor as an independent field. This removes the derivative dependence from the conformal rescaling itself; the original metric construction is recovered by imposing a separate constraint on the auxiliary field.

Introduce an independent auxiliary scalar $X$ and define
\begin{equation}
 \gE_{\mu\nu}=F(X)g_{\mu\nu}.
 \label{eq:generic_parent_weyl_X}
\end{equation}

This map is algebraic on $(g_{\mu\nu},X)$. The original metric sector is selected by
\begin{equation}
 X=R[g],
 \label{eq:generic_parent_projection_X}
\end{equation}
or, after eliminating $g_{\mu\nu}$,
\begin{equation}
 \hPF[\gE,X]\equiv R[F(X)^{-1}\gE]-X=0.
 \label{eq:generic_PF_def}
\end{equation}
Thus, in the localized description, a projected configuration is simply a pair $(\gE_{\mu\nu},X)$ satisfying Eq.~\eqref{eq:generic_PF_def}. Equivalently, the auxiliary carrier $X$ equals the curvature of the reconstructed metric $g_{\mu\nu}=F(X)^{-1}\gE_{\mu\nu}$. Explicitly $\hPF[\gE,X]$ is given by
\begin{equation}
\hPF=F\RE+3F_X\BoxE X
 +3\left(F_{XX}-\frac32\frac{F_X^2}{F}\right)(\nablaE X)^2-X.
 \label{eq:generic_PF_explicit}
\end{equation}
Defining
\begin{equation}
 \mathcal A(X)=F_{XX}-\frac32\frac{F_X^2}{F},
 \label{eq:generic_Q_aux}
\end{equation}
let
\begin{equation}
 \chi\equiv\delta X
\end{equation}
denote an infinitesimal variation in the auxiliary-scalar direction. The scalar-direction linearization is then
\begin{align}
 \cN_F\chi
 ={}&3F_X\BoxE\chi
 +6 \mathcal A\,\nablaE^\mu X\,\nablaE_\mu\chi
 \nonumber\\
 &+\left(
 F_X\RE-1
 +3F_{XX}\BoxE X
 +3\mathcal{A}_{X}(\nablaE X)^2
 \right)\chi .
 \label{eq:generic_NF_operator}
\end{align}
In particular,
\begin{equation}
 \delta\!\left[3\mathcal{A}(X)(\nablaE X)^2\right]=3\mathcal{A}_X(\nablaE X)^2\chi+6\mathcal{A}\,\nablaE^\mu X\,\nablaE_\mu\chi .
\end{equation}

On a constant projected background, the first-derivative term proportional to $\mathcal{A}$ in Eq~\eqref{eq:generic_NF_operator} does not affect the principal symbol and vanishes yielding
\begin{equation}
 \cN_F=3F_X\BoxE+\frac{XF_X}{F}-1\; ,
 \label{eq:generic_NF_constant}
\end{equation}
and the homogeneous algebraic diagnostic is
\begin{equation}
 F-XF_X=0\; .
 \label{eq:generic_auxiliary_fold}
\end{equation}
Actual failure of the graph requires noninvertibility of the chosen operator realization, not merely this algebraic condition.

Now consider the full class
\begin{equation}
 \widetilde g_{\mu\nu}=F(R[g])\, g_{\mu\nu},
 \qquad F>0,
 \qquad F_R\neq0.
 \label{eq:general_R_dependent_conformal_map}
\end{equation}
The localized projection is given by Eq.~\eqref{eq:generic_PF_def} with linearization
\begin{equation}
 \delta\widehat{\mathcal P}_F=B_F[\gamma]+\mathcal N_F\chi.
 \label{eq:general_R_linearized_projection}
\end{equation}

To test whether the inverse can be a finite-order differential operator, it is sufficient to examine its highest-derivative momentum dependence. This is encoded in the principal symbol. At a point $x$, freeze the smooth background coefficients and retain the highest-derivative terms. Let $\xi_\mu\in T_x^*\mathcal M$ denote the local momentum covector used in the symbol calculation, with $\xi^2\equiv\widetilde g^{\mu\nu}\xi_\mu\xi_\nu$; equivalently, locally one replaces $\widetilde\nabla_\mu\mapsto i\xi_\mu$. The principal symbol is obtained by keeping only the highest-derivative terms and making this replacement. It therefore describes the leading short-wavelength response of the differential operator. For a noncharacteristic covector, meaning a momentum direction for which the highest-derivative part of the scalar operator is nonzero---here $\xi^2\neq0$---the principal symbols are
\begin{align}
 \sigma_2(\mathcal N_F)(x,\xi)
 &=-3F_X(x)\xi^2,
 \label{eq:general_R_normal_symbol}\\
 \sigma_2(B_F)(x,\xi)\,\gamma
 &=F(x)\left(
 -\xi^\mu\xi^\nu\gamma_{\mu\nu}
 +\xi^2\gamma
 \right).
 \label{eq:general_R_source_symbol}
\end{align}
Consequently,
\begin{multline}
 \sigma_0(-\mathcal N_F^{-1}B_F)(x,\xi)\!\cdot\!\gamma
 =-\frac{F}{3F_X}
 \left(\frac{\xi^\mu\xi^\nu}{\xi^2}\gamma_{\mu\nu}-\gamma\right).
 \label{eq:general_R_complete_response_symbol}
\end{multline}
The factor $\xi^\mu\xi^\nu/\xi^2$ selects the component of a metric perturbation aligned with the momentum covector $\xi_\mu$; this is the sense in which it acts as a longitudinal projector. The essential feature for the present argument is the denominator $1/\xi^2$. A finite-order differential operator has polynomial momentum dependence, whereas the inverse response in Eq.~\eqref{eq:general_R_complete_response_symbol} contains an inverse power of momentum. It therefore cannot be represented by finitely many local derivatives. In Lorentzian signature, the set $\xi^2=0$ is the characteristic set of the wave operator. Across this set a global inverse is specified by a Green prescription, such as a retarded or advanced inverse, together with the corresponding boundary or Cauchy data. Away from the characteristic set the inverse may be described microlocally---that is, locally in both spacetime position and momentum---as a pseudodifferential operator, a class of operators that allows nonpolynomial momentum dependence such as $1/\xi^2$.

The metric inverse itself follows from
\begin{equation}
 \delta g_{\mu\nu}=F^{-1}\gamma_{\mu\nu}
 -F^{-2}F_X\chi\,\widetilde g_{\mu\nu}.
 \label{eq:general_inverse_metric_variation}
\end{equation}
After imposing $X=R[g]$, so that $F_X$ and $F_R$ denote the derivative
of the same branch function at the projected curvature,
\begin{align}
 \sigma_0(\delta g_{\mu\nu})(x,\xi)
 ={}&\frac{1}{F}\gamma_{\mu\nu}
 \nonumber\\
 &+\frac{1}{3F}
 \left(\frac{\xi^\alpha\xi^\beta}{\xi^2}\gamma_{\alpha\beta}-\gamma\right)
 \widetilde g_{\mu\nu}.
 \label{eq:general_inverse_metric_symbol}
\end{align}
This obstruction is not confined to pure diffeomorphism directions; for
example, the constant-background trace sector below retains a nonpolynomial
factor when the scalaron mass is nonzero.

The preceding argument can be formalized as follows. Let $F\in C^3$, with $F>0$ and $F_R\neq0$, and suppose the map in Eq.~\eqref{eq:general_R_dependent_conformal_map} admitted a $C^1$, finite-jet inverse on an open set of unrestricted metrics, where ``unrestricted'' means that no field equations, symmetry reduction, or special restriction on the metric perturbations has been imposed. The linearization of that inverse would be a finite-order differential operator and hence have a polynomial symbol. Equation~\eqref{eq:general_inverse_metric_symbol} is nonpolynomial, giving a contradiction. Thus branchwise functional inverses may exist after differential data are fixed, but no generic finite-jet metric-only inverse exists.

This result has a direct methodological consequence. Curvature-dependent conformal transformations of the class considered here should not generically be treated as ordinary local changes of metric variables. Their use as forward maps of known configurations remains valid, but inverse and off-shell constructions must retain the differential projection together with the functional domain and boundary or Cauchy data required to solve it. In particular, a curvature-dependent conformal rule used in modified gravity must be distinguished according to whether it defines a forward map, a local transformation on an enlarged parent field space, or a genuine change of off-shell variables.

This distinction is essential for variational applications. Keeping $R$ explicit in the conformal factor does not remove the constraint, since $R=R[g]$ is not an independent field. The transformed variables must still satisfy the differential projection~\eqref{eq:generic_PF_def}, and their admissible variations therefore obey Eq.~\eqref{eq:general_R_linearized_projection}. In particular, fixing the curvature carrier, $\chi=0$, restricts the transformed metric variation to $B_F[\gamma]=0$ rather than allowing an arbitrary $\gamma_{\mu\nu}$. Consequently, naively treating $\widetilde g_{\mu\nu}$ as an unconstrained variational variable at fixed curvature does not reproduce the variational problem of the original metric theory, as will be shown explicitly in Secs.~\ref{sec:fr_parent}--\ref{sec:hessian}. Conversely, declaring the curvature carrier independent enlarges the field space, in which case the differential projection must be retained as a separate constraint. The corresponding boundary or Cauchy data must likewise be carried consistently through the transformation.

For factors depending on other curvature invariants, the same procedure produces a normal-operator matrix. Its complete composite response, rather than the differential order of the normal matrix alone, must be tested in each model.

\subsection{Off-shell fibers and frame copies}
\label{subsec:off_shell_fibers}

A fixed transformed metric can have several off-shell preimages. Define
\begin{equation}
 \cF_{\gE}=\{g_{\mu\nu}:F(R[g])g_{\mu\nu}=\gE_{\mu\nu}\}.
 \label{eq:fiber_def}
\end{equation}
Thus the fiber $\cF_{\gE}$ is simply the set of all original metrics that are mapped to the same transformed metric $\gE_{\mu\nu}$. If this set contains more than one element, the transformed metric alone does not select a unique preimage.

On a locally monotonic branch, let $r_F$ denote the local inverse function
of $F$, and write
\begin{equation}
 g_{\mu\nu}=\varphi^2\gE_{\mu\nu},
 \qquad r_F=F^{-1}.
 \label{eq:fiber_phi_def}
\end{equation}
Then
\begin{equation}
 R[g]=r_F(\varphi^{-2}),
 \label{eq:fiber_R_condition}
\end{equation}
and
\begin{equation}
 R[\varphi^2\gE]
 =\varphi^{-2}\left(\RE-6\varphi^{-1}\BoxE\varphi\right)
 \label{eq:R_phi_relation}
\end{equation}
yield
\begin{equation}
 6\BoxE\varphi
 =\varphi\left[\RE-\varphi^2r_F(\varphi^{-2})\right].
 \label{eq:fiber_equation}
\end{equation}
Thus a fiber is itself a differential solution space.

For $F(R)=1+2\alpha R$, $\alpha>0$, and
 $\gE_{\mu\nu}=\eta_{\mu\nu}$, homogeneous positive-branch fibers obey
\begin{equation}
 \ddot\varphi+\frac{\varphi(\varphi^2-1)}{12\alpha}=0,
 \label{eq:fiber_oscillator}
\end{equation}
with
\begin{equation}
 E_\varphi=\frac12\dot\varphi^2
 +\frac{(\varphi^2-1)^2}{48\alpha}.
 \label{eq:fiber_energy}
\end{equation}
Every orbit satisfying
\begin{equation}
 0<E_\varphi<\frac{1}{48\alpha}
 \label{eq:positive_fiber_bound}
\end{equation}
oscillates around $\varphi=1$ without crossing zero. Hence the same transformed Minkowski metric has infinitely many regular off-shell preimages. Most are not solutions of the metric theory, so this multiplicity does not conflict with on-shell equivalence; it shows that a global metric-only chart requires a section prescription.

\section{Metric \texorpdfstring{$f(R)$}{f(R)} gravity and its local parent}
\label{sec:fr_parent}

\subsection{Metric action and Legendre branch}

Consider metric $f(R)$ gravity with minimally coupled matter,
\begin{equation}
 S_f[g,\psi]=\frac{1}{2\kapp^2}
 \int\dd^4x\sqrt{-g}\,f(R[g])+S_m[g,\psi].
 \label{eq:Sf_full}
\end{equation}
Its field equation and trace are
\begin{equation}
 f_RR_{\mu\nu}-\frac12fg_{\mu\nu}
 +(g_{\mu\nu}\Box-\nabla_\mu\nabla_\nu)f_R
 =\kapp^2T_{\mu\nu},
 \label{eq:metric_fr_eom}
\end{equation}
\begin{equation}
 3\Box f_R+f_RR-2f=\kapp^2T.
 \label{eq:trace_fr}
\end{equation}
The trace displays the scalaron, the propagating spin-zero degree of freedom of the higher-derivative metric theory \cite{Sotiriou:2008rp,Dolgov:2003px,DeFelice:2010aj}.

On a regular Legendre branch introduce $X$ and
\begin{equation}
 \Phi=f_X(X),\qquad f_{XX}\neq0,\qquad\Phi>0,
 \label{eq:Legendre_def}
\end{equation}
with inverse $X=X(\Phi)$. The Legendre potential is
\begin{equation}
 U(\Phi)=\Phi X(\Phi)-f(X(\Phi)),
 \qquad U_\Phi=X(\Phi),
 \label{eq:U_def}
\end{equation}
and the Jordan-frame parent action is
\begin{equation}
 S_J=\frac{1}{2\kapp^2}\int\dd^4x\sqrt{-g}\,[\Phi R-U(\Phi)]
 +S_m[g,\psi].
 \label{eq:SJ_full}
\end{equation}
Varying $\Phi$ gives
\begin{equation}
 R=X(\Phi).
 \label{eq:aux_eq_J}
\end{equation}
Substitution recovers Eq.~\eqref{eq:Sf_full}. This standard auxiliary representation \cite{Whitt:1984pd,Maeda:1988ab,Teyssandier:1983zz,Wands:1993uu, Magnano:1993bd} enlarges the off-shell configuration space: $X$ agrees with the geometric curvature only on the constraint. The condition $f_{XX}\neq0$ makes the Legendre map locally invertible, while $\Phi>0$ ensures that the conformal rescaling $\widetilde g_{\mu\nu}=\Phi g_{\mu\nu}$ preserves the metric signature and gives the Einstein--Hilbert term its standard sign. The quantity $U_{\Phi\Phi}=1/f_{XX}$ is not by itself the canonically normalized scalaron mass.

\subsection{Einstein-frame parent and projection}

Define
\begin{equation}
 \gE_{\mu\nu}=\Phi g_{\mu\nu},
 \qquad s=\ln\Phi.
 \label{eq:EF_def}
\end{equation}
Because $\Phi$ is independent, this is a local parent-field
redefinition. Up to the standard boundary completion,
\begin{align}
 S_E[\gE,s,\psi]
 ={}&\frac{1}{2\kapp^2}\int\dd^4x\sqrt{-\gE}
 \left[\RE-\frac32(\nablaE s)^2-W(s)\right]
 \nonumber\\
 &+S_m[e^{-s}\gE,\psi],
 \label{eq:SE_full}
\end{align}
where
\begin{equation}
 W(s)=e^{-2s}U(e^s).
 \label{eq:W_def}
\end{equation}
The scalar Euler derivative is defined by
\begin{equation}
 \delta_sS_E=\langle E_s,\delta s\rangle,
 \qquad
 E_s=\frac{2\kapp^2}{\sqrt{-\gE}}\frac{\delta S_E}{\delta s},
 \label{eq:Es_def_full}
\end{equation}
with scalar pairing
\begin{equation}
 \langle A,B\rangle
 \equiv\frac{1}{2\kapp^2}\int\dd^4x\sqrt{-\gE}\,AB.
 \label{eq:scalar_pairing_def}
\end{equation}
For minimally coupled matter the scalar equation has the schematic form
\begin{equation}
 3\BoxE s-W_s+
 \frac{2\kapp^2}{\sqrt{-\gE}}\frac{\delta S_m[e^{-s}\gE,\psi]}{\delta s}=0.
 \label{eq:parent_scalar_eom_generic}
\end{equation}

The original metric theory is recovered by imposing
\begin{equation}
 \hPf[\gE,s]\equiv
 e^s\left[\RE+3\BoxE s-\frac32(\nablaE s)^2\right]-X(e^s)=0.
 \label{eq:Pf_full}
\end{equation}
We will call a configuration ``projected'' when it satisfies Eq.~\eqref{eq:Pf_full}, similarly to the previous definition. This condition identifies the independent parent scalar with the curvature degree of freedom of the original metric theory. A projected configuration need not satisfy the parent or metric field equations; when both the projection and the dynamical equations hold, we will refer explicitly to a common-shell configuration.

It is therefore represented by the differential constraint set
\begin{equation}
 \cM_f=\{(\gE_{\mu\nu},s):\hPf[\gE,s]=0\}
 \label{eq:Mf_def}
\end{equation}
inside the parent space. To represent this constraint set locally as a graph over the Einstein-frame metric, the partial linearization of \(\hPf\) in the scalar direction must be an isomorphism between the chosen function spaces. Under this hypothesis, the implicit-function theorem gives a locally unique functional \(s_\star[\gE]\), so that
\begin{equation}
 \iota_{G_P}:\gE_{\mu\nu}\mapsto
 (\gE_{\mu\nu},s_\star[\gE]),
 \qquad \hPf[\gE,s_\star[\gE]]=0.
 \label{eq:section_iota_def}
\end{equation}
A ``section'' is a prescription that selects one allowed parent configuration above each retained metric. A ``graph section'' means, more specifically, that this choice can be written as a single-valued functional $s=s_\star[\gE]$ on the selected domain. Thus, once the functional domain and boundary or Cauchy data are fixed, each admissible $\gE_{\mu\nu}$ is assigned one scalar configuration.

The scalaron is not an extra fundamental field coordinate of the original metric action, but it remains its genuine physical spin-zero mode. The parent scalar represents that mode locally; in metric-only Einstein-side variables it is reconstructed through the differential section and its homogeneous
data.

\begin{figure}[t]
 \centering
 \begin{tikzpicture}[
 >=Stealth,
 every node/.style={font=\normalsize}
 ]
 
 \node (JP) at (0,0) {$(g_{\mu\nu},\Phi)$};
 \node (EP) at (6.4,0) {$(\widetilde g_{\mu\nu},\Phi)$};
 \node (JG) at (0,-2.8) {$g_{\mu\nu}$};
 \node (EG) at (6.4,-2.8) {$\widetilde g_{\mu\nu}\ \text{alone}$};
 
 \draw[->]
 (JP) -- node[above] {$\mathcal F_P:\ \widetilde g_{\mu\nu}=\Phi g_{\mu\nu}$}
 (EP);
 
 \draw[->]
 (JP) -- node[right] {$R[g]=X(\Phi)$}
 (JG);
 
 \draw[->]
 (EP) -- node[left] {\ $R[\Phi^{-1}\widetilde g]=X(\Phi)$}
 (EG);
 
 \draw[line width=0.8pt]
 ($(EP)!0.55!(EG)+(-0.16,0.16)$)
 --
 ($(EP)!0.55!(EG)+(0.16,-0.16)$);
 \draw[line width=0.8pt]
 ($(EP)!0.55!(EG)+(-0.16,-0.16)$)
 --
 ($(EP)!0.55!(EG)+(0.16,0.16)$);
 
 \draw[->]
 (JG) -- node[above] {\small not a pointwise metric map}
 (EG);
 
 \draw[line width=0.8pt]
 ($(JG)!0.55!(EG)+(-0.16,0.16)$)
 --
 ($(JG)!0.55!(EG)+(0.16,-0.16)$);
 \draw[line width=0.8pt]
 ($(JG)!0.55!(EG)+(-0.16,-0.16)$)
 --
 ($(JG)!0.55!(EG)+(0.16,0.16)$);
 
 \end{tikzpicture}
 \caption{Projection to the geometric metric $f(R)$ sector and conformal transformation do not commute as pointwise operations. The upper arrow is algebraic because it is performed on the parent field space. The lower arrow is not a pointwise map because the conformal factor is $f_R(R[g])$.}
 \label{fig:noncomm_diagram}
\end{figure}
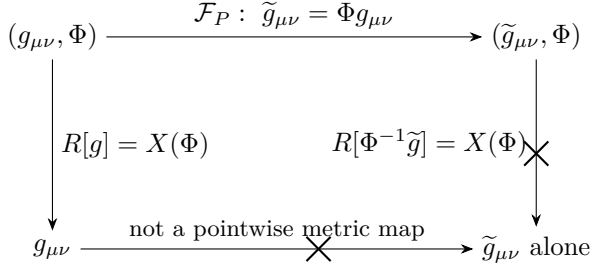

This distinction is summarized in Fig.~\ref{fig:noncomm_diagram}. The upper horizontal arrow is a local and algebraically invertible Weyl transformation on the enlarged parent field space, $(g_{\mu\nu},\Phi)\mapsto(\widetilde g_{\mu\nu},\Phi)$, with $\widetilde g_{\mu\nu}=\Phi g_{\mu\nu}$. The vertical arrows impose the geometric projection $R[g]=X(\Phi)$. On the Einstein-frame side, this
projection becomes
\begin{equation}
 R[\Phi^{-1}\widetilde g]=X(\Phi),
\end{equation}
which is differential in $\Phi$. Consequently, the lower horizontal relation between $g_{\mu\nu}$ and $\widetilde g_{\mu\nu}$ is not an independent pointwise change of metric variables: it is defined only after solving the differential projection with a specified branch and appropriate boundary or Cauchy data.

\subsection{Parent-shell identity and forward images}
\label{subsec:parent_shell_projection_identity}

Exact parent solutions must be distinguished from arbitrary parent configurations. In vacuum, the parent metric trace and scalar equations are
\begin{equation}
 \RE-\frac32(\nablaE s)^2-2W=0,
 \label{eq:parent_trace_vac_identity}
\end{equation}
\begin{equation}
 3\BoxE s-W_s=0.
 \label{eq:parent_scalar_vac_identity}
\end{equation}
Using
\begin{equation}
 W_s=e^{-s}X(e^s)-2W,
 \label{eq:Ws_identity_X}
\end{equation}
one finds
\begin{align}
 \hPf
 =e^s\bigg[&\RE-\frac32(\nablaE s)^2-2W
 +3\BoxE s-W_s\bigg].
 \label{eq:projection_parent_shell_identity}
\end{align}
Thus $\hPf=0$ follows from the full parent equations; matter terms cancel in the same combination. Exact regular parent and metric solutions therefore coincide on the regular branch. Off shell, however, arbitrary parent configurations generally violate the projection.

A forward image of a known solution is also different from a closed variational formulation in transformed variables. The former uses the known preimage and its curvature. The latter must define an action and fluctuation domain without that information. The local parent does so on $(\widetilde g_{\mu\nu},s)$; a metric-only Einstein-side formulation does so only after the differential section and its functional data have been specified.

\section{Normal operator and inverse sections}
\label{sec:normal_operator}

\subsection{Linearized projection and the normal operator}

Let
\begin{equation}
 \gamma_{\mu\nu}=\delta\gE_{\mu\nu},
 \qquad \sigma=\delta s .
 \label{eq:variation_gamma_sigma}
\end{equation}
The variation of the projection is
\begin{equation}
 \delta\hPf=B[\gamma]+\cL_\Phi\sigma,
 \label{eq:linear_projection_def}
\end{equation}
Here $B[\gamma]$ is the change of the projection produced by varying the metric while holding $s$ fixed. By the ``vertical'' variation we mean the opposite operation: varying $s$ while holding the retained metric $\gE_{\mu\nu}$ fixed. The operator $\cL_\Phi$ governing this scalar variation will be called the ``normal operator''. The terminology refers only to the metric--scalar splitting used to eliminate $s$; no notion of orthogonality in configuration space is assumed. Imposing $\hPf=0$, one finds
\begin{equation}
 \cL_\Phi
 =3\Phi\left(\BoxE-\nablaE^\mu s\,\nablaE_\mu\right)
 +X(\Phi)-\Phi X_\Phi(\Phi),
 \qquad \Phi=\ee^s .
 \label{eq:Lphi_full}
\end{equation}
A direct derivation of Eq.~\eqref{eq:Lphi_full}, including the use of the projection condition $\Phi Q=X(\Phi)$, is given in Appendix~\ref{app:normal_operator}. The term $3\Phi\BoxE$ is the principal part on a genuine $f(R)$ branch with principal part being the highest-derivative part of the operator that controls its characteristic and short-wavelength behavior. Since
\begin{equation}
 X_\Phi=\frac{1}{f_{RR}},
 \label{eq:Xphi_fRR}
\end{equation}
this operator is second order when $f_{RR}\neq0$.

Allowed tangent fluctuations to the metric $f(R)$ section obey
\begin{equation}
 B[\gamma]+\cL_\Phi\sigma=0 .
 \label{eq:tangent_full}
\end{equation}
On a regular functional domain for which the selected boundary or Cauchy
problem admits a right inverse $G_P$, the general solution can be written
as
\begin{equation}
 \sigma=-G_PB[\gamma]+\sigma_h,
 \qquad
 \cL_\Phi\sigma_h=0 .
 \label{eq:tangent_solution_full}
\end{equation}
A right inverse $G_P$ constructs one particular solution of the sourced equation: applying $\cL_\Phi$ to that solution reproduces the source $-B[\gamma]$. The term $\sigma_h$ solves the corresponding homogeneous equation and contains the complementary boundary or initial data that are not determined by the source. If those data are part of the definition of the section, they also fix $\sigma_h$. In a globally hyperbolic Lorentzian problem the normally hyperbolic part of $\cL_\Phi$ admits advanced and retarded Green operators under the usual hypotheses; the existence of homogeneous wave solutions does not by itself obstruct a retarded or advanced inverse with fixed Cauchy data \cite{Bar:2013bpw,Salgado:2005hx}.

When the selected operator has a nontrivial kernel, an
ordinary two-sided inverse does not exist. A generalized inverse may then
be introduced on complementary subspaces, schematically,
\begin{equation}
 \cL_\Phi G_P=I-\Pi_{\rm coker},
 \qquad
 G_P\cL_\Phi=I-\Pi_{\rm ker},
 \label{eq:generalized_inverse_projectors}
\end{equation}
where $\Pi_{\rm ker}$ projects onto $\ker\cL_\Phi$ and $\Pi_{\rm coker}$ onto the obstruction space, identified with $\ker\cL_\Phi^\dagger$ for a Fredholm realization\footnote{A Fredholm realization is a boundary-value problem for which the kernel and cokernel are finite dimensional and the image is closed, allowing failures of invertibility to be characterized by these finite-dimensional spaces~\cite{BandaraGoffengSaratchandran2023}.}. The kernel of $\cL_\Phi$ consists of scalar perturbations that the operator maps to zero. The cokernel measures source directions that are not in the image of $\cL_\Phi$; for a Fredholm realization it can be identified with the kernel of the adjoint operator. The tangent equation is then solvable only if
\begin{equation}
 \Pi_{\rm coker}B[\gamma]=0 .
 \label{eq:tangent_fredholm_condition}
\end{equation}
Thus, the metric-induced source must have no component along an adjoint zero mode, otherwise the scalar projection equation has no solution in the chosen functional domain. Homogeneous solutions of the projection equation and zero modes of a specified boundary-value operator must not be conflated. The former are complementary data of the linearized projection problem. On a common-shell physical background, after the coupled field equations and constraints have been imposed, they coincide with scalaron Cauchy data. The latter obstruct the use of $\widetilde g_{\mu\nu}$ as a unique graph coordinate for that particular functional problem.

On a constant projected background the normal operator has Klein--Gordon form with
\begin{equation}
 m_E^2
 =
 \frac{f_R-Rf_{RR}}{3f_Rf_{RR}},
 \qquad
 m_J^2=f_Rm_E^2.
 \label{eq:mass_relation_EJ}
\end{equation}
Here $m_E$ is the Einstein-metric normal-response parameter and $m_J$ is its Jordan-metric counterpart. When the background also satisfies the common field equations, they describe the same physical scalaron pole after the conformal rescaling of momenta. Off shell they should be interpreted as parameters of the linearized projection and trace operators, rather than as an independently observable particle mass. They are not numerically equal unless $f_R=1$ that corresponds to GR.

The same differential operator follows directly from the intrinsic Jordan-frame metric equations. To see this, we will define the trace residual
\begin{equation}
  \mathcal E_J[g,T]
  \equiv
  3\Box f_R+f_R R-2f-\kapp^2T .
  \label{eq:Jordan_trace_residual}
\end{equation}
On a constant-curvature configuration, its first variation is
\begin{equation}
  \delta\mathcal E_J=3f_{RR}\left(\Box-m_J^2\right)\delta R-\kapp^2\delta T,
  \label{eq:Jordan_trace_Frechet}
\end{equation}
where
\begin{equation}
  m_J^2=
  \frac{f_R-Rf_{RR}}{3f_{RR}} .
  \label{eq:Jordan_scalar_pole_crosscheck}
\end{equation}
Thus the same response operator is defined even away from a solution. If the background also satisfies the field equations, \(\delta\mathcal E_J=0\) gives the usual linearized trace equation,
\begin{equation}
  \left(\Box-m_J^2\right)\delta R=\frac{\kapp^2}{3f_{RR}}\delta T .
\end{equation}
On the same background,
\begin{equation}
 \sigma=\delta\ln f_R=\frac{f_{RR}}{f_R}\delta R,
 \qquad
 \Box=f_R\BoxE .
 \label{eq:Jordan_Einstein_scalar_relation}
\end{equation}
The homogeneous Jordan-frame operator is therefore proportional to
\[
 3f_R^2\left(\BoxE-m_E^2\right)\sigma
 =
 f_R\,\mathcal L_\Phi\sigma .
\]
Hence the Jordan-frame trace operator, and the Einstein-frame normal
operator have the same kernel and the same pole parameter after the constant
conformal rescaling, $m_J^2=f_Rm_E^2$. On a common-shell background this
is the physical scalaron pole; off shell it is the corresponding response
pole of the two linearized operators. This provides an intrinsic
metric-theory cross-check of the normal-operator interpretation.

The linearized projection condition~\eqref{eq:tangent_full} does not eliminate the scalaron as a propagating degree of freedom. It expresses the fact that, in the original metric theory, the scalaron is not an additional fundamental field coordinate independent of the metric. It is the spin-zero component of the higher-derivative metric sector. In metric-only Einstein variables this mode is reconstructed through the chosen inverse of $\cL_\Phi$, together with the allowed homogeneous data. The finite-jet obstruction therefore sharpens the status of the scalaron: it is intrinsic to the metric theory, but it cannot generically be represented as a local finite-order functional of the Einstein metric alone.

Equation~\eqref{eq:tangent_solution_full} is the infinitesimal form of the
fixed-point obstruction. If $s_\star[\gE]$ were a finite-jet functional,
then $\delta s_\star$ would be a finite-order differential operator acting
on $\gamma_{\mu\nu}$. Instead, on a regular section, its sourced first
variation is obtained by applying the selected Green operator to
 $B[\gamma]$, while the homogeneous part is fixed separately by the chosen
boundary or Cauchy data. The complete nonlinear section still requires
solving $\hPf=0$ and controlling its branch, domain, global existence,
uniqueness, and possible bifurcations. A metric-only Einstein-side
formulation is therefore incomplete until these data and the treatment of any
kernel have been specified.

\subsection{Metric variation \texorpdfstring{$B[\gamma]$}{B}}

For later use we record the explicit form of $B[\gamma]$. Let
\begin{equation}
 \mathcal R_s[\gE]
 =\ee^s\left[
 \RE+3\BoxE s-\frac32(\nablaE s)^2
 \right].
 \label{eq:Rs_def}
\end{equation}
At fixed $s$,
\begin{equation}
 B[\gamma]=\delta_{\gE}\mathcal R_s[\gamma].
 \label{eq:B_def}
\end{equation}
Using
\begin{equation}
 \delta\RE
 =-\RE^{\mu\nu}\gamma_{\mu\nu}
 +\nablaE^\mu\nablaE^\nu\gamma_{\mu\nu}
 -\BoxE\gamma,
 \qquad
 \gamma=\gE^{\mu\nu}\gamma_{\mu\nu},
 \label{eq:delta_R_tilde}
\end{equation}
and the fixed-scalar variation of $\BoxE s$, one obtains
\begin{align}
 B[\gamma]
 &=\ee^s\Big[
 -\RE^{\mu\nu}\gamma_{\mu\nu}
 +\nablaE^\mu\nablaE^\nu\gamma_{\mu\nu}
 -\BoxE\gamma
 \nonumber\\
 &\quad
 -3\gamma^{\mu\nu}\nablaE_\mu\nablaE_\nu s
 -3\left(\nablaE_\mu\gamma^{\mu\nu}-\frac12\nablaE^\nu\gamma\right)\nablaE_\nu s
 \nonumber\\
 &\quad
 +\frac32\gamma^{\mu\nu}\nablaE_\mu s\nablaE_\nu s
 \Big],
 \label{eq:B_explicit}
\end{align}
where indices on $\gamma_{\mu\nu}$ are raised with $\gE^{\mu\nu}$.

The generic arbitrary-background symbol proof has already been given in Sec.~\ref{sec:curvature_maps}. Here we use a ``frozen symbol'', obtained by evaluating the background coefficients at one spacetime point and treating them as constants while examining the momentum dependence. This is a local diagnostic and becomes an exact ordinary Fourier multiplier only on a translationally invariant background. On a general constant- $s$ background this is the frozen symbol at a spacetime point. It becomes an exact ordinary Fourier multiplier only on a flat translationally invariant background. On a nonflat maximally symmetric background, the corresponding exact global statement must instead be formulated through an appropriate harmonic or spectral decomposition.

For $\widetilde\nabla_\mu s=0$, use
 $\widetilde\nabla_\mu\mapsto i\xi_\mu$, and define
 $\ell_\Phi(\xi)$ and $b(\xi)$ by
\[
 \widehat{\mathcal L_\Phi\sigma}(\xi)
 =-\ell_\Phi(\xi)\widehat\sigma(\xi),
 \qquad
 \widehat{B[\gamma]}(\xi)
 =-b(\xi)\!\cdot\!\widehat\gamma(\xi).
\]
The full frozen symbols are then
\begin{align}
 \ell_\Phi(\xi)
 &=3\Phi\bigl(\xi^2+m_E^2\bigr),
 \label{eq:L_symbol_constant_background}
 \\
 b(\xi)\!\cdot\!\widehat\gamma
 &=\Phi\left(
  \widetilde R^{\mu\nu}\widehat\gamma_{\mu\nu}
  +\xi^\mu\xi^\nu\widehat\gamma_{\mu\nu}
  -\xi^2\widehat\gamma
 \right).
 \label{eq:B_symbol_constant_background}
\end{align}
Consequently,
\begin{equation}
 \widehat{\delta s_\star}(\xi)
 =-\frac{
  \widetilde R^{\mu\nu}\widehat\gamma_{\mu\nu}
  +\xi^\mu\xi^\nu\widehat\gamma_{\mu\nu}
  -\xi^2\widehat\gamma
 }{3(\xi^2+m_E^2)}.
 \label{eq:complete_tangent_symbol}
\end{equation}

For a pure-trace perturbation,
 $\widehat\gamma_{\mu\nu}
=\tfrac14\widehat\gamma\,\widetilde g_{\mu\nu}$, this reduces to
\begin{equation}
 \widehat{\delta s_\star}(\xi)
 =-\frac{\widetilde R-3\xi^2}
 {12(\xi^2+m_E^2)}\,\widehat\gamma .
 \label{eq:pure_trace_complete_tangent_symbol}
\end{equation}
Using
\begin{equation}
 \widetilde R=\frac{R}{f_R},
 \qquad
 3m_E^2
 =\frac{f_R-Rf_{RR}}{f_Rf_{RR}},
\end{equation}
one obtains the identity
\begin{equation}
 \widetilde R+3m_E^2=\frac{1}{f_{RR}}.
 \label{eq:trace_residue_identity}
\end{equation}
Equation~\eqref{eq:pure_trace_complete_tangent_symbol} may therefore be
rewritten as
\begin{equation}
 \widehat{\delta s_\star}(\xi)
 =
 \left[
  \frac14
  -\frac{1}{12f_{RR}}
   \frac{1}{\xi^2+m_E^2}
 \right]\widehat\gamma .
 \label{eq:pure_trace_complete_tangent_decomposition}
\end{equation}
On a regular Legendre branch, $f_{RR}$ is finite and nonzero, so the residue of the inverse differential operator cannot vanish. Hence the pure-trace response is nonpolynomial on every regular constant projected background, including when $m_E^2=0$. A generic metric perturbation also retains the nonpolynomial longitudinal structure $\xi^\mu\xi^\nu/\xi^2$. Thus the finite-jet obstruction is not confined to diffeomorphism-gauge directions.

Physically, $B[\gamma]$ is the source induced by a metric fluctuation $\gamma_{\mu\nu}$ in the linearized transformed auxiliary relation $R[\ee^{-s}\widetilde g]=X(\ee^s)$. On the complete parent shell, this relation can be written as a combination of the Einstein-frame metric trace and scalar equations, but off shell it should not be identified with the Jordan-frame trace equation itself. Accordingly, $B[\gamma]$ contains not only the standard Einstein-frame Ricci variation $-\widetilde{R}^{\mu\nu}\gamma_{\mu\nu}+\widetilde{\nabla}^\mu\widetilde{\nabla}^\nu\gamma_{\mu\nu} -\widetilde{\Box}\gamma$, but also derivative couplings to the background scalar $s$. The terms proportional to $\nablaE_\mu s$ encode how a varying scalar profile mixes metric and scalar perturbations. In the projected tangent lift $\sigma=-G_P B[\gamma]$, this source entirely determines the scalar response to a given metric deformation, up to homogeneous zero modes. The expression is therefore an essential building block for cosmological perturbation theory and for the projected Hessian computed in later sections.

\section{Boundary conditions and variational data}
\label{sec:boundary_terms}

\subsection{Boundary completion}

A differential expression does not define an inverse until its functional domain and boundary or Cauchy data are fixed. The same principle applies to the variational problem. For metric $f(R)$ gravity one commonly adds
\begin{equation}
 S_{\partial f}=\frac{1}{\kapp^2}
 \int_{\partial\cM}\dd^3x\sqrt{|h|}\,f_RK,
 \label{eq:fR_boundary}
\end{equation}
where $|h|$ is the determinant of the induced metric on the boundary, while fixing the induced metric and an additional higher-derivative datum, for example $\delta R|_{\partial\cM}=0$, equivalently $\delta f_R|_{\partial\cM}=0$, on a regular branch \cite{Dyer:2008hb,Guarnizo:2010xr}. Mixed and Hamiltonian prescriptions are also possible; the important point is that the boundary term alone does not remove the need to control the additional metric data.

The Jordan scalar--tensor parent has the corresponding completion
\begin{equation}
 S_{\partial J}=\frac{1}{\kapp^2}
 \int_{\partial\cM}\dd^3x\sqrt{|h|}\,\Phi K.
 \label{eq:parent_boundary_J}
\end{equation}
Under the local parent Weyl transformation, the normal-derivative part of
 $\Phi K$ combines with the boundary contribution generated by the
transformed bulk curvature. The result is the standard Einstein-frame
Gibbons--Hawking--York term with consistently transformed scalar boundary
data; there is no universal uncancelled scalar boundary term. Dirichlet data
for the Einstein metric and the independent parent scalar are a standard
second-order prescription and are not generically overdetermining.

\subsection{Operator realization}

The differential operator $\cL_\Phi$ does not by itself determine
 $G_P$. One must specify the source and solution spaces, boundary or initial
conditions, and the treatment of homogeneous and zero modes. These data are
part of the section
\begin{equation}
 \iota_{G_P}:\gE_{\mu\nu}\longmapsto
 (\gE_{\mu\nu},s_\star[\gE]).
 \label{eq:section_with_domain_data}
\end{equation}
A Green operator fixes the sourced solution; it is not equivalent to choosing
a homogeneous mode. In Lorentzian signature, retarded and advanced operators
are causal solution maps. In Euclidean signature, Dirichlet, Neumann, Robin,
or mixed conditions define different elliptic realizations and spectra.
These choices need not describe different physical theories when they are
mapped consistently from the same Jordan-frame problem.

Consequently, the pulled-back Hessian, static scalar charge, cosmological
response, and nonlocal metric kernel inherit the inverse prescription of the
problem under study. A variational principle does not by itself choose a
retarded, advanced, Feynman, or Euclidean inverse. The metric-only
Einstein-side formulation is therefore incomplete unless the functional data
defining its differential section are stated explicitly.

\section{Projected Hessian and constrained Gaussian theory}
\label{sec:hessian}

Having fixed the functional domain and inverse prescription defining a
regular section, we can now differentiate that section and construct the
metric-theory Hessian. At a spectral obstruction the following formulae
apply only after projection to a complementary subspace and imposition of
the compatibility condition~\eqref{eq:tangent_fredholm_condition}.

\subsection{Second variation of the section}

We define the pulled-back metric-only Einstein-side action
\begin{equation}
 S_\star[\gE]=S_E[\gE,s_\star[\gE]],
 \qquad
 \hPf[\gE,s_\star[\gE]]=0 .
 \label{eq:Sstar_def}
\end{equation}
The word ``pullback'' means that the parent action is evaluated only on the constraint section by replacing the independent scalar with the functional $s_\star[\gE]$. The resulting action therefore depends on the metric alone.

For two metric fluctuations $\gamma_1,\gamma_2$, let the corresponding tangent lifts be \footnote{A ``tangent lift'' is the combined metric--scalar perturbation that remains tangent to the constraint set. Its scalar component is therefore not chosen independently: $\sigma_a$ is the change in $s_\star$ required by the metric perturbation $\gamma_a$.}
\begin{equation}
 \eta_a=(\gamma_a,\sigma_a),
 \qquad
 \sigma_a=\delta s_\star[\gamma_a],
 \qquad a=1,2 .
 \label{eq:tangent_lifts_two}
\end{equation}

Differentiating the projection a second time gives
\begin{equation}
 \cL_\Phi\,\delta^2s_\star[\gamma_1,\gamma_2]+\delta^2\hPf[\eta_1,\eta_2]=0 .
 \label{eq:second_projection_full}
\end{equation}
On a regular section, with the same inverse prescription used in the tangent
lift,
\begin{equation}
 \delta^2s_\star[\gamma_1,\gamma_2]=-G_P\,\delta^2\hPf[\eta_1,\eta_2],
 \label{eq:second_section_full}
\end{equation}
where the second-order homogeneous data are fixed by the definition of the
section. If $\cL_\Phi$ has a kernel, the right-hand side must satisfy the
corresponding adjoint-kernel compatibility condition and the section need
not be a graph in $\widetilde g_{\mu\nu}$ alone.

The second variation of the pulled-back action is
\begin{equation}
 \delta^2S_\star[\gamma_1,\gamma_2]=\delta^2S_E[\eta_1,\eta_2]+\left\langle E_s,\delta^2s_\star[\gamma_1,\gamma_2]\right\rangle .
 \label{eq:Hchain_full}
\end{equation}
Substituting Eq.~\eqref{eq:second_section_full} yields the projected Hessian
\begin{equation}
 \delta^2S_\star[\gamma_1,\gamma_2]=\delta^2S_E[\eta_1,\eta_2]-\left\langle E_s, G_P\,\delta^2\hPf[\eta_1,\eta_2] \right\rangle .
 \label{eq:Hsplit_full}
\end{equation}
The second term will be called the ``embedding correction''. It appears because the constraint section is generally curved inside the larger metric--scalar field space: even when a perturbation remains on the section at first order, a further scalar adjustment is required at second order.

A compact derivation of Eqs.~\eqref{eq:second_projection_full}--\eqref{eq:Hsplit_full} in condensed field-space notation is given in Appendix~\ref{app:second_variation}. Equation~\eqref{eq:Hsplit_full} is one of the main results of the paper.

The projected Hessian separates the classical second variation of the metric $f(R)$ theory into two contributions. The first term, $\delta^2S_E[\eta_1,\eta_2]$, is the Hessian of the scalar-tensor parent evaluated on tangent vectors that already satisfy the linearized projection. It contains the standard Einstein-frame graviton and scalar kinetic terms, and it coincides with what one would obtain by naively inserting the tangent lift into the parent action. The second term is the \emph{embedding correction} arising from the second derivative of the selected section inside the parent field space. This correction is proportional to the parent scalar Euler derivative $E_s$. It therefore vanishes on the parent scalar shell $E_s=0$, and hence in particular on the full parent shell. For special fluctuation directions it may also vanish accidentally away from the scalar shell. On a generic off-shell background, however, it contributes to the difference between the metric-theory Hessian and the corresponding parent quadratic form. In geometric language, the embedding term is the field-theoretic analogue of the second fundamental form of the constraint surface $\hPf=0$. It encodes how the scalar fluctuation is forced to adjust at second order to keep the configuration on the projected constraint set.

It is useful to isolate the general geometric content. Suppose a parent field
space with coordinates $(q,u)$ is constrained by $P(q,u)=0$, and assume
that the normal derivative $P_u$ admits an inverse $G_P$ on a chosen
functional domain. The section $u_\star[q]$ satisfies
\begin{equation}
 \delta u_\star=-G_P P_q\delta q,
 \quad
 \delta^2u_\star=-G_P\delta^2P[(\delta q,\delta u),(\delta q,\delta u)] .
 \label{eq:general_section_identity}
\end{equation}
For any parent action $S(q,u)$, the pulled-back Hessian is
\begin{equation}
 \delta^2(S\circ\iota)
 =
 \delta^2S\big|_{T\cM_P}
 +
 \left\langle S_u,\delta^2u_\star\right\rangle .
 \label{eq:general_pullback_hessian}
\end{equation}
Equation~\eqref{eq:Hsplit_full} is the specialization of
Eq.~\eqref{eq:general_pullback_hessian} to the metric $f(R)$ projection.
The last term is equation-of-motion proportional in the normal direction. It
therefore vanishes on the full parent shell, but it is part of the off-shell
metric-theory Hessian.

This structure implies that a semiclassical
calculation intended to represent the original metric theory in Einstein
variables must use both the projected fluctuation domain and the complete
pulled-back Hessian. Omitting either ingredient fails to reproduce the
metric-theory Gaussian problem. Omitting both yields the unrestricted
scalar--tensor parent, whereas retaining the tangent restriction but dropping
the embedding term gives an incomplete restricted-parent quadratic form.
This conclusion does not eliminate the physical scalaron; it prevents that
mode from being counted again as an arbitrary parent-scalar fluctuation at
fixed projected metric. A complete one-loop calculation must supplement the
projected Hessian with the projected gauge fixing, ghost operator, pushforward
measure, constraint Jacobian, regularization, and zero-mode prescription.

\subsection{Projected versus unrestricted Gaussian theories}

Equation~\eqref{eq:Hsplit_full} is the projected Einstein-frame Hessian of the metric theory. Its first term is the scalar-tensor parent Hessian restricted to projected tangent lifts. Its second term is the embedding contribution of the section $\hPf=0$. This term is absent if $(\gE_{\mu\nu},s)$ are quantized as unrestricted parent variables, but it is required when the metric $f(R)$ theory is represented inside the Einstein-frame parent.

In Jordan variables no analogous term is displayed, because the theory is already intrinsic to the metric field space:
\begin{equation}
 \delta^2 S_f[g][h_1,h_2]
 \equiv
 H^{\rm JF}_{f(R)}[h_1,h_2].
 \label{eq:jordan_intrinsic_hessian}
\end{equation}
There is no additional scalar field coordinate to project out, because the physical scalaron is already encoded in the higher-derivative metric fluctuation. In Einstein variables, however, the same metric theory is obtained only after choosing a section $s_\star[\gE]$ satisfying $\hPf[\gE,s_\star[\gE]]=0$. The corresponding Hessian is therefore section-dependent:
\begin{equation}
 H^{\rm proj}_{\rm EF}[G_P]
 \neq
 H^{\rm parent}_{\rm EF}
 \qquad
 {\rm off\ shell}.
 \label{eq:projected_not_parent_hessian}
\end{equation}
More explicitly, the three quadratic objects that must not be conflated are
\begin{align}
H_{\rm metric}^{\rm proj}
&=H_{\rm parent}\big|_{T\mathcal M_f}+H_{\rm embed},
\label{eq:three_hessians_projected}
\\
H_{\rm restricted}
&=H_{\rm parent}\big|_{T\mathcal M_f},
\label{eq:three_hessians_restricted}
\\
H_{\rm parent}
&=H_{\rm parent}[(\gamma_{\mu\nu},\sigma)
\text{ arbitrary}].
\label{eq:three_hessians_parent}
\end{align}
The second line is obtained by imposing the tangent relation but omitting the embedding term; it is neither the metric-theory Hessian nor the unrestricted parent Hessian. The inequivalence appears before any determinant is evaluated. The unrestricted parent calculation misses the $f(R)$ projection in two ways: it integrates over arbitrary $(\gamma_{\mu\nu},\sigma)$, rather than projected lifts satisfying Eq.~\eqref{eq:tangent_full}, and it omits the section-curvature term in Eq.~\eqref{eq:Hsplit_full}.

Accordingly, after a projected gauge fixing has been specified, the metric-theory one-loop functional in Einstein variables has the schematic structure
\begin{align}
\Gamma^{(1)}_{{\rm metric},\,{\rm EF}}
={}&
\frac12\Tr_{T\cM_f}\log H^{\rm proj}_{{\rm EF},{\rm gf}}[G_P]-\Tr\log\cM^{\rm proj}_{\rm FP}
\nonumber\\
&+
\Gamma^{(1)}_{\rm ind}[G_P]
+\cdots ,
\label{eq:one_loop_projected_full}
\end{align}
where $\cM^{\rm proj}_{\rm FP}$ is the Faddeev--Popov operator of the pulled-back gauge condition and $\Gamma^{(1)}_{\rm ind}$ denotes the contribution of the induced measure and of the constraint Jacobian.

The unrestricted scalar--tensor parent instead gives
\begin{align}
\Gamma^{(1)}_{{\rm parent},\,{\rm EF}}
={}&
\frac12
\Tr_{\rm parent}
\log H^{\rm parent}_{{\rm EF},{\rm gf}}
-
\Tr\log\cM^{\rm parent}_{\rm FP}
\nonumber\\
&+
\Gamma^{(1)}_{{\rm parent},{\rm meas}}
+\cdots .
\label{eq:one_loop_parent_full}
\end{align}
Thus Eq.~\eqref{eq:Hsplit_full} does not by itself prescribe every determinant entering a complete one-loop calculation. It identifies the field-space origin of the different Gaussian kernels appearing in the known on-shell/off-shell comparisons
\cite{RufSteinwachs2018a,Ruf:2017xon,Ohta:2017trn,
Falls:2018olk}.

The distinction between Eqs.~\eqref{eq:one_loop_projected_full} and \eqref{eq:one_loop_parent_full} is structural rather than a discrepancy generated only by regularization. The traces act on different fluctuation domains, and the corresponding quadratic kernels already differ before a determinant, heat-kernel coefficient, or gauge-parameter limit is evaluated. Consequently, the unrestricted-parent result cannot in general be converted into the metric-theory result by merely appending or deleting a scalar determinant after the fact. The tangent restriction, Green prescription, second-order embedding contribution, projected gauge sector, and induced measure must be implemented consistently at the level of the constrained functional integral.

\subsection{Constrained functional integral and induced measure}

A schematic constrained representation of the original metric path integral in Einstein variables is
\begin{align}
Z_{f,{\rm EF}}
={}&\int\mathcal D\gE\,\mathcal D s\,
\mathcal J_{\rm push}[\gE,s]\,
\Delta^{\rm proj}_{\rm FP}\,
\delta\!\left[\hPf[\gE,s]\right]
\nonumber\\
&\times
\exp\!\left\{i\left(S_E+S^{\rm proj}_{\rm gf}\right)\right\},
\label{eq:constrained_path_integral}
\end{align}
where $\mathcal J_{\rm push}$ denotes the pushforward of the chosen Jordan-frame metric measure, including any normal Jacobian. The ``pushforward measure'' is the functional measure obtained when the original Jordan-frame metric integration variables are rewritten in the Einstein-frame parent variables. Accordingly, $\mathcal J_{\rm push}$ denotes the Jacobian factors associated with this change of variables, including the normal direction selected by the constraint~\cite{Falls:2018olk,Kuntz:2026vhs}. This factor is not optional: an arbitrary flat measure in $(\widetilde g,s)$ would define a different quantum theory. The delta functional can equivalently be represented by a Lagrange multiplier,
\begin{equation}
\delta[\hPf]
\propto
\int\mathcal D\lambda\,
\exp\!\left(i\langle\lambda,\hPf\rangle\right).
\label{eq:constraint_multiplier_representation}
\end{equation}
This is the functional-integral version of retaining the auxiliary constraint emphasized in recent analyses of off-shell equivalence \cite{Kuntz:2026vhs}.

On a regular section, the linearized delta functional imposes
\begin{equation}
 B[\gamma]+\cL_\Phi\sigma=0,
\end{equation}
and integration over the normal fluctuation produces both the tangent restriction $\sigma=-G_PB[\gamma]$ and the determinant associated with the chosen realization of $\cL_\Phi$. Its precise power, phase, and adjoint structure depend on the fundamental measure, signature, gauge fixing, and boundary conditions. If $\cL_\Phi$ has a kernel, the delta functional additionally enforces the cokernel compatibility condition and the allowed kernel coefficients remain variables whose weight must be read from the complete constrained action Hessian. A normal zero mode is not \emph{generically} enough to infer a flat direction. On the constant parent-shell quadratic example of Sec.~\ref{sec:nonlocal_action}, however, $\cL_\Phi$ is proportional to the scalar vertical Hessian, and its kernel is flat at Gaussian order; that special case is treated explicitly below.

On the full parent shell $E_s=0$, the embedding term vanishes, so this particular source of off-shell mismatch disappears, consistently with the established on-shell agreement. Off shell, the projected metric Hessian contains the section-curvature contribution absent from the unrestricted parent Hessian. The present construction therefore fixes the correct fluctuation domain and quadratic kernel, but a numerical one-loop effective action still requires a definite measure, regularization, gauge, boundary, and zero-mode prescription.

\subsection{Geometric interpretation and configuration-space covariance}

Equation~\eqref{eq:Hsplit_full} is the field-theory version of the ordinary chain rule for a function restricted to a constraint set. The first term is the parent Hessian evaluated on projected tangent lifts; the second contracts the normal Euler derivative with the second derivative of the section. In the finite-dimensional analogue,
\begin{equation}
 \frac{\dd^2}{\dd x^2}S(x,y_\star(x))
 =S_{xx}+2S_{xy}y_\star'+S_{yy}(y_\star')^2+S_yy_\star'',
 \label{eq:finite_dim_analogy}
\end{equation}
and the last term is precisely the embedding contribution. If a configuration-space metric and connection are introduced, this statement has the covariant form
\begin{equation}
 \mathfrak D_i\mathfrak D_j(S_E\circ\iota)
 =e_i{}^Ae_j{}^B\nabla_A\nabla_BS_E+K_{ij}{}^A\nabla_AS_E,
 \label{eq:covariant_pullback_hessian}
\end{equation}
where $K_{ij}{}^A$ is the normal component of the covariant derivative of the tangent basis. This connects the exact coordinate identity used here to the Vilkovisky--DeWitt construction without requiring a particular configuration-space geometry in the main analysis \cite{Vilkovisky:1984st,DeWitt:2003pm,Kuntz:2026vhs}.

\subsection{Explicit off-shell correction in \texorpdfstring{$R+\alpha R^2$}{R+alpha R2}}
\label{subsec:explicit_offshell_hessian_example}

The embedding term can be made completely explicit on a simple off-shell
background. Let
\begin{equation}
 f(R)=R+\alpha R^2,
 \qquad
 X(\Phi)=\frac{\Phi-1}{2\alpha} .
 \label{eq:explicit_staro_hessian_model}
\end{equation}
Choose a constant projected background with $s=s_0$, $\Phi=\ee^{s_0}$, and an Einstein metric satisfying
\begin{equation}
 \RE=\frac{X(\Phi)}{\Phi}
 =\frac{\Phi-1}{2\alpha\Phi} .
 \label{eq:explicit_staro_projected_background}
\end{equation}
It lies on $\hPf=0$, but it is not on the parent scalar shell unless
\begin{equation}
 W_s=\frac{\Phi-1}{2\alpha\Phi^2}=0 .
 \label{eq:explicit_staro_Ws}
\end{equation}
On this background
\begin{equation}
 \cL_\Phi=3\Phi\BoxE-\frac{1}{2\alpha},
 \qquad
 G_P=\left(3\Phi\BoxE-\frac{1}{2\alpha}\right)^{-1} .
 \label{eq:explicit_staro_LG}
\end{equation}
Consider the conformal metric variation
\begin{equation}
 \gamma_{\mu\nu}=2\chi\gE_{\mu\nu} .
 \label{eq:explicit_conformal_mode}
\end{equation}
Then
\begin{equation}
 B[\gamma]=-\Phi(2\RE\chi+6\BoxE\chi),
 \qquad
\sigma_\chi=G_P\Phi(2\RE\chi+6\BoxE\chi) .
 \label{eq:explicit_B_sigma_chi}
\end{equation}
Using exponential conformal coordinates,
\begin{equation}
 \gE_{\mu\nu}(\epsilon)=\ee^{2\epsilon\chi}\bar\gE_{\mu\nu},
 \qquad
 s(\epsilon)=\bar s+\epsilon\sigma,
 \label{eq:explicit_exp_coords}
\end{equation}
the conformal-sector second variation of the projection is
\begin{align}
 \delta^2\hP_{f,\chi}={}&\Phi\Big[
 \RE(\sigma-2\chi)^2
 \nonumber\\
 &+2(\sigma-2\chi)
 (-6\BoxE\chi+3\BoxE\sigma)
 \nonumber\\
 &-12(\nablaE\chi)^2
 +12\nablaE\chi\!\cdot\!\nablaE\sigma
 \nonumber\\
 &-3(\nablaE\sigma)^2
 \Big]
 -\frac{\Phi}{2\alpha}\sigma^2 .
 \label{eq:explicit_d2P_chi}
\end{align}
The correction to the pulled-back quadratic form is therefore
\begin{equation}
 \Delta H_\chi
 =-\frac{1}{2\kapp^2}
 \int\dd^4x\sqrt{-\gE}\,
 E_s\,G_P\,\delta^2\hP_{f,\chi}[\sigma_\chi] .
 \label{eq:explicit_DeltaH_formal}
\end{equation}
For a compact Euclidean constant-curvature background and a constant conformal mode $\chi=\chi_0$, with $G_P$ acting on the constant subspace,
\begin{equation}
 \sigma_{\chi_0}=-2(\Phi-1)\chi_0,
 \qquad
 \delta^2\hP_{f,\chi_0}
 =\frac{2\Phi(\Phi-1)}{\alpha}\chi_0^2 .
 \label{eq:explicit_constant_mode_projection}
\end{equation}
Since $E_s=-W_s$ on the constant background, one obtains, in the exponential conformal coordinates of Eq.~\eqref{eq:explicit_exp_coords},
\begin{equation}
 \Delta H_{\chi_0}^{(\mathrm{exp})}
 =
 -\frac{V_E}{\kapp^2}
 \frac{(\Phi-1)^2}{\alpha\Phi}\chi_0^2,
 \qquad
 V_E=\int\dd^4x\sqrt{\gE},
 \label{eq:explicit_DeltaH_number}
\end{equation}
up to the conventional Lorentzian-to-Euclidean overall sign. In this coordinate choice, the embedding contribution is nonzero for $\Phi\neq1$ and vanishes when the parent scalar equation is imposed.

Equation~\eqref{eq:explicit_DeltaH_number} is the ordinary off-shell second-variation contribution in the exponential field coordinates chosen in Eq.~\eqref{eq:explicit_exp_coords}. Its isolated numerical coefficient is not invariant under a nonlinear reparametrization of the metric configuration space: ordinary off-shell Hessians, and their decomposition into restricted and embedding terms, change by contributions proportional to the first Euler derivatives. The parametrization-covariant statement is instead the covariant pullback identity \eqref{eq:covariant_pullback_hessian}, in which the normal gradient is contracted with the second fundamental form of the selected section.

The calculation therefore establishes the nonvanishing of the embedding contribution in the stated exponential parametrization and illustrates its vanishing on the common shell. Its sign or isolated coefficient should not by itself be interpreted as a parametrization-independent stability criterion. Such a conclusion would require the complete covariant, gauge-fixed Euclidean Hessian, together with the Wick-rotation convention, the conformal-factor contour, and all metric components. For nonconstant perturbations, the same coordinate representation contains the Green operator $G_P$ and is consequently nonlocal.
 
The formal construction of the projected fluctuation space and its quadratic operator is now complete. We next show that the same normal Green operator has directly observable manifestations. We begin with the static weak-field response of compact sources and then turn to homogeneous and perturbed cosmology.

\section{Weak-field limit and compact scalar charges}
\label{sec:weakfield}

The same differential denominator that appears in the normal operator also appears, on the common weak-field background, in the scalaron trace response. The scalar charge is fixed only after the metric field equations, source, regularity, matching, and asymptotic boundary conditions are imposed. Weak-field and post-Newtonian limits of analytic metric $f(R)$ models have long exhibited the same massive scalar pole and Yukawa response \cite{Capozziello:2007ms,Capozziello:2009vr, Stabile:2010zk,Berry:2011pb}.

\subsection{Flat \texorpdfstring{$R+\alpha R^2$}{R+alpha R2} branch}

Consider the quadratic model \cite{Starobinsky:1980te}
\begin{equation}
 f(R)=R+\alpha R^2,
 \qquad
 \alpha>0,
 \label{eq:staro_branch}
\end{equation}
for which
\begin{equation}
 \Phi=1+2\alpha R,
 \qquad
 X(\Phi)=\frac{\Phi-1}{2\alpha},
 \qquad
 m_s^2=\frac{1}{6\alpha}.
 \label{eq:staro_defs}
\end{equation}
The exact projection is
\begin{equation}
 \ee^s\left[
 \RE+3\BoxE s-\frac32(\nablaE s)^2
 \right]
 =\frac{\ee^s-1}{2\alpha}.
 \label{eq:staro_projection}
\end{equation}
About the flat projected branch
 $\widetilde g_{\mu\nu}=\eta_{\mu\nu}$, $s=0$,
\begin{equation}
 \cL_\Phi=3(\BoxE-m_s^2),
 \label{eq:L_staro_flat_lorentz}
\end{equation}
and in the static sector
\begin{equation}
 \cL_\Phi=3(\Delta_E-m_s^2).
 \label{eq:L_staro_static}
\end{equation}
The full linearized projection is
\begin{equation}
 \widetilde R^{(1)}[\gamma]
 +3(\BoxE-m_s^2)s=0.
 \label{eq:staro_projection_linear_flat}
\end{equation}
Thus $(\BoxE-m_s^2)s=0$ describes only a vertical homogeneous perturbation at fixed flat Einstein metric; the sourced tangent relation contains the linearized metric curvature in Eq.~\eqref{eq:staro_projection_linear_flat}.

\subsection{Green representation and scalar charge}

The Jordan-frame trace equation is
\begin{equation}
 6\alpha\Box R_J-R_J=\kapp^2T,
 \label{eq:trace_staro}
\end{equation}
or
\begin{equation}
 (\Box-m_s^2)R_J=\frac{\kapp^2}{6\alpha}T.
 \label{eq:trace_staro_mass}
\end{equation}
On the flat weak-field branch the two metrics coincide at zeroth order and
\begin{equation}
 s=\ln(1+2\alpha R_J)
 =2\alpha R_J+\cO(R_J^2).
 \label{eq:s_R_linear}
\end{equation}
For a nonrelativistic source, $T\simeq-\rho$, the projected static scalar therefore obeys
\begin{equation}
 (\Delta_E-m_s^2)s=-\frac{\kapp^2}{3}\rho.
 \label{eq:static_scalar_source}
\end{equation}
The normal inverse is, up to the overall factor of three,
\begin{equation}
 G_P^{\rm stat}
 =\frac13(\Delta_E-m_s^2)^{-1},
 \label{eq:Gp_static_staro}
\end{equation}
and the decaying elliptic Yukawa Green function is defined by
\begin{equation}
 (\Delta_E-m_s^2)G_m(\vec x,\vec x')
 =-\delta^{(3)}(\vec x-\vec x'),
\end{equation}
\begin{equation}
 G_m(\vec x,\vec x')
 =\frac{\ee^{-m_s|\vec x-\vec x'|}}
 {4\pi|\vec x-\vec x'|}.
 \label{eq:Yukawa_green}
\end{equation}
Consequently,
\begin{equation}
 s(\vec x)
 =\frac{\kapp^2}{3}
 \int\dd^3x'\,G_m(\vec x,\vec x')\rho(\vec x').
 \label{eq:s_green_static}
\end{equation}
For a spherical body of radius $R_b$, regularity at the origin, matching at the surface, and decay at infinity give
\begin{equation}
 s_{\rm ext}(r)
 =\frac{Q_s\ee^{-m_sr}}{r},
 \qquad r>R_b,
 \label{eq:s_ext_Q}
\end{equation}
with
\begin{equation}
 Q_s
 =\frac{\kapp^2}{3m_s}
 \int_0^{R_b}\dd\bar r\,
 \bar r\sinh(m_s\bar r)\rho(\bar r).
 \label{eq:Qs_formula}
\end{equation}
This is the classical response associated with the same scalaron denominator that appears in the normal operator. In the regular static, asymptotically decaying section, $Q_s$ is fixed by the source and boundary data rather than being an additional free exterior parameter. A free incoming or outgoing scalaron wave is not forbidden by the metric theory; it belongs to a different Lorentzian Cauchy problem with different homogeneous data. It is excluded only by the static regularity and decay conditions used to define the section above.

\subsection{Metric potentials}

In Jordan-frame Newtonian gauge,
\begin{equation}
 \dd s_J^2=-(1+2\Psi)\dd t^2
 +(1-2\Phi_N)\delta_{ij}\dd x^i\dd x^j,
 \label{eq:weak_metric_J}
\end{equation}
the point-mass solution is
\begin{align}
 \Psi(r)&=-\frac{GM}{r}\left(1+\frac13\ee^{-m_sr}\right),
 \\
 \Phi_N(r)&=-\frac{GM}{r}\left(1-\frac13\ee^{-m_sr}\right).
 \label{eq:yukawa_potentials}
\end{align}
For a point source Eq.~\eqref{eq:Qs_formula} gives $Q_s=\kapp^2M/(12\pi)$. The same scalaron pole therefore appears both in the normal inverse and in the standard $1/3$ Yukawa correction. Screening in viable nonlinear models changes the effective charge through the background-dependent mass and matching problem, but it does not remove the scalaron degree of freedom \cite{Khoury:2003aq,Khoury:2003rn,
HuSawicki2007}.

\section{Cosmological projection and scalaron response}
\label{sec:cosmology}
In cosmology the same obstruction becomes an initial-value problem. A prescribed Einstein-frame expansion history does not determine the curvature-dependent conformal factor algebraically; the projected scalar must instead satisfy a nonlinear differential section equation with independent branch and Cauchy data.

\subsection{FLRW variables in the two frames}

A spatially flat Einstein-frame FLRW metric is
\begin{equation}
 \dd\widetilde s^2=-\dd\widetilde t^2+\widetilde a(\widetilde t)^2\delta_{ij}\dd x^i\dd x^j,
 \qquad
 \widetilde H=\frac{1}{\widetilde a}\frac{\dd\widetilde a}{\dd\widetilde t} .
 \label{eq:EF_FLRW_metric}
\end{equation}
If $g_{\mu\nu}=\ee^{-s}\gE_{\mu\nu}$, then the Jordan cosmic time and scale factor are
\begin{equation}
 \dd t_J=\ee^{-s/2}\dd\widetilde t,
 \qquad
 a_J=\ee^{-s/2}\widetilde a .
 \label{eq:frame_scale_time}
\end{equation}
The Jordan Hubble parameter is therefore
\begin{equation}
 H_J
 =\ee^{s/2}\left(\widetilde H-\frac12\dot s\right),
 \label{eq:HJ_HE_s}
\end{equation}
where a dot denotes $\dd/\dd\widetilde t$. The Einstein-frame Ricci scalar is
\begin{equation}
 \RE=6(\dot{\widetilde H}+2\widetilde H^2).
 \label{eq:RE_FLRW}
\end{equation}
For homogeneous $s(\widetilde t)$,
\begin{equation}
 \BoxE s=-\ddot s-3\widetilde H\dot s,
 \qquad
 (\nablaE s)^2=-\dot s^2 .
 \label{eq:boxs_FLRW}
\end{equation}
Therefore the projection equation becomes the nonlinear second-order ODE
\begin{equation}
 \ee^s\left[
 6(\dot{\widetilde H}+2\widetilde H^2)
 -3\ddot s-9\widetilde H\dot s+\frac32\dot s^2
 \right]
 =X(\ee^s).
 \label{eq:FLRW_projection_ODE}
\end{equation}
This equation is the cosmological version of the fixed-point problem. Given $\widetilde a(\widetilde t)$, the scalar $s$ is not determined algebraically. It requires initial data and a branch prescription.

In the cosmological context, Eq.~\eqref{eq:FLRW_projection_ODE} is a nonlinear consistency condition that ties the parent coordinate $s$ to a prescribed Einstein-frame expansion history. It is not an additional field equation to be imposed on top of the complete parent system. Locally, a second-order projection equation admits two pieces of initial data when a regular branch exists, but the physically relevant history is fixed only after the branch and the Jordan-equivalent Cauchy or boundary data are specified. These homogeneous data belong to the inverse projection problem and are not generated by a Green operator. On a common-shell cosmological solution they coincide with the scalaron Cauchy data of the metric theory. A Green operator enters only after linearization, where it constructs the particular response to a change of the prescribed metric history.

\subsection{Homogeneous normal operator}

For homogeneous variations $\omega=\delta s$ at fixed $\gE$, Eq.~\eqref{eq:Lphi_full} gives
\begin{equation}
 \cL_\Phi\omega
 =3\ee^s\left[-\ddot\omega-(3\widetilde H-\dot s)\dot\omega\right]
 +\left[X-\ee^sX_\Phi\right]\omega .
 \label{eq:L_FLRW_hom}
\end{equation}
Equivalently,
\begin{equation}
 \cL_\Phi\omega
 =-3\ee^s\left[\ddot\omega+(3\widetilde H-\dot s)\dot\omega+M_{\rm eff}^2\omega\right],
 \label{eq:L_FLRW_effmass_form}
\end{equation}
where
\begin{equation}
 M_{\rm eff}^2
 =\frac{\ee^sX_\Phi-X}{3\ee^s}.
 \label{eq:M_eff_FLRW}
\end{equation}
The operator has the form of a damped oscillator with friction coefficient $3\widetilde H-\dot s$ and normal-response parameter $M_{\rm eff}^2$. On a time-dependent background the sign of this coefficient alone is not a complete diagnostic of the behavior of the projection equation: anti-damping can occur if $3\widetilde H-\dot s<0$, and all coefficients may vary in time.

On a constant- $s$, constant-curvature projected background one has that the mass is given by Eq.~\eqref{eq:mass_relation_EJ}. At this stage $M_{\rm eff}^2$ is a parameter of the normal response. If the background also satisfies the common metric--parent field equations, then it coincides with the Einstein-frame physical scalaron mass parameter. On such a common-shell background, and for $f_R>0$, positivity requires
\begin{equation}
 \frac{f_R-Rf_{RR}}{f_{RR}}>0 .
 \label{eq:mass_positivity_condition}
\end{equation}
On the standard viable branch $f_{RR}>0$, this reduces to $f_R-Rf_{RR}>0$, consistently with the standard scalaron stability requirements and the absence of the Dolgov--Kawasaki matter instability \cite{Dolgov:2003px,Faraoni:2006sy,DeFelice:2010aj}. Away from the common shell, however, the same quantity should be interpreted as a normal-response coefficient rather than as an independently observable particle mass or, by itself, a physical stability criterion. Its sign is also logically distinct from invertibility of a particular boundary-value realization of $\cL_\Phi$.

This expression diagnoses the homogeneous scalar normal sector and appears again in the discussion of graph breakdown. Stability of the complete cosmological solution additionally requires the coupled metric constraints, tensor and matter perturbations, kinetic and gradient matrices, gauge conditions, and background evolution.

\subsection{Starobinsky cosmological projection}

For $f(R)=R+\alpha R^2$, Eq.~\eqref{eq:FLRW_projection_ODE} becomes
\begin{equation}
 \ee^s\left[
 6(\dot{\widetilde H}+2\widetilde H^2)
 -3\ddot s-9\widetilde H\dot s+\frac32\dot s^2
 \right]
 =\frac{\ee^s-1}{2\alpha}.
 \label{eq:FLRW_staro_projection}
\end{equation}
In the regime where both the scalar amplitude $s$ and the Einstein-frame curvature $\RE$ are small, the nonlinear terms can be neglected and the equation reduces to a driven linear oscillator.
For weak scalar amplitude, this gives
\begin{equation}
 \ddot s+3\widetilde H\dot s+m_s^2s
 \simeq
 2(\dot{\widetilde H}+2\widetilde H^2)
 =\frac{\RE}{3}.
 \label{eq:FLRW_staro_linear}
\end{equation}
The left-hand side is the damped scalaron operator with the canonical mass $m_s^2=1/(6\alpha)$; the right-hand side is the Einstein-frame Ricci curvature, acting as a source. This equation is the linearised projection constraint: it forces the scalaron to respond to the expansion history. The solution is
\begin{equation}
 s(\widetilde t)
 =s_h(\widetilde t)+\int\dd\widetilde t'\,G_{\rm ret}(\widetilde t,\widetilde t')\frac{\RE(\widetilde t')}{3},
 \label{eq:FLRW_staro_green}
\end{equation}
where $G_{\rm ret}$ is the retarded solution operator of the damped scalaron equation and $s_h$ is fixed by the chosen initial data. This is a causal solution map: the response at $\widetilde t$ depends on the earlier curvature history. The homogeneous solution $s_h$ is ordinary scalaron Cauchy data, not automatically a zero mode obstructing inversion. Moreover, a retarded solution operator should not be identified directly with the symmetric kernel of an ordinary single-history nonlocal action; that distinction is developed in Sec.~\ref{sec:EFT_comments}. This is the cosmological analogue of the compact-source scalar charge: the particular, curvature-sourced response is the retarded Green image of the normal operator, whereas $s_h$ is independent complementary Cauchy data.

The intrinsic Jordan-frame Friedmann equations provide the standard metric-theory background evolution, while the Einstein-frame parent treats $s$ as an independent scalar coupled to $\ee^{-s}\gE_{\mu\nu}$. The projected Einstein cosmology is obtained only after imposing Eq.~\eqref{eq:FLRW_projection_ODE}; equivalently, the curvature reconstructed from $\ee^{-s}\gE$ must equal $X(\ee^s)$ throughout the evolution. Hence the Jordan-frame reconstruction equation and the Einstein-side section equation describe the same metric-theory background from two complementary directions, but only the latter displays explicitly the branch data and Green prescription required to reconstruct $s$ from a prescribed Einstein-frame expansion history.
\subsection{Scalar perturbations and the quasistatic limit}

Consider scalar perturbations in the Einstein-frame metric,
\begin{equation}
 \dd\widetilde s^2
 =-(1+2\widetilde\Psi)\dd\widetilde t^2
 +\widetilde a^2(1-2\widetilde\Phi_N)\delta_{ij}\dd x^i\dd x^j,
 \label{eq:EF_scalar_pert_metric}
\end{equation}
and write
\begin{equation}
 s(\widetilde t,\vec x)=\bar s(\widetilde t)+\delta s(\widetilde t,\vec x).
 \label{eq:s_pert}
\end{equation}
The projected fluctuation is not arbitrary. In Fourier space it obeys
\begin{equation}
 \cL_\Phi(k)\,\delta s_k
 =-B_k[\widetilde\Phi_N,\widetilde\Psi],
 \label{eq:cosmo_tangent_fourier}
\end{equation}
where $\cL_\Phi(k)$ is obtained from Eq.~\eqref{eq:Lphi_full} by replacing the spatial Laplacian with $-k^2/\widetilde a^2$. The projection is geometric and contains no separate matter source. Matter enters indirectly because the metric perturbations appearing in $B_k$ satisfy the matter-sourced metric field equations. Equation~\eqref{eq:cosmo_tangent_fourier} does not remove the physical scalaron perturbation. It expresses that, once the projected metric perturbation and the homogeneous section data are specified, there is no additional freely adjustable parent-scalar fluctuation at fixed metric.

In the quasistatic subhorizon regime, the operator is dominated by
\begin{equation}
 \cL_\Phi(k)\simeq -3\Phi\left(\frac{k^2}{\widetilde a^2}+m_E^2\right),
 \label{eq:L_quasistatic}
\end{equation}
where
\begin{equation}
 m_E^2=\frac{f_R-Rf_{RR}}{3f_Rf_{RR}}
 \label{eq:ms_cosmo}
\end{equation}
is the Einstein-frame mass parameter on a slowly varying background. Thus
\begin{equation}
 \delta s_k\sim
 \frac{1}{3\Phi}
 \frac{B_k}{k^2/\widetilde a^2+m_E^2}.
 \label{eq:delta_s_quasistatic}
\end{equation}

\paragraph*{Observable quasistatic closure.}

In the Jordan frame, using the Newtonian gauge metric \eqref{eq:weak_metric_J} on an expanding background, the quasistatic subhorizon limit gives the well-known effective Newton coupling and slip. Define
\begin{equation}
 Q(k,a)=\frac{k^2}{a^2}\frac{f_{RR}}{f_R}.
 \label{eq:Q_pert_def}
\end{equation}
Then, for nonrelativistic matter and slowly varying backgrounds, one obtains
\begin{align}
 \frac{k^2}{a^2}\Psi
 &\simeq -4\pi G_{\rm eff}(k,a)\rho\delta,
 \label{eq:modified_poisson}
 \\
 \frac{G_{\rm eff}}{G}
 &=\frac{1}{f_R}\frac{1+4Q}{1+3Q},
 \label{eq:Geff_fR}
 \\
 \frac{\Phi_N}{\Psi}
 &\simeq\frac{1+2Q}{1+4Q}.
 \label{eq:slip_fR}
\end{align}
These expressions are standard consequences of metric $f(R)$ gravity \cite{Song:2006ej,DeFelice:2010aj}. Their scalar-sector denominator is closely related to the normal-operator denominator, but the two are not exactly identical before the background-curvature truncation used in the quasistatic approximation. Indeed, with $a\equiv a_J$,
\begin{align}
 1+3Q
 &=
 \frac{3f_{RR}}{f_R}
 \left(
  \frac{k^2}{a_J^2}
  +\frac{f_R}{3f_{RR}}
 \right)
 \nonumber\\
 &=
 \frac{3f_{RR}}{f_R}
 \left(
  \frac{k^2}{a_J^2}
  +m_J^2
  +\frac{R}{3}
 \right),
 \label{eq:quasistatic_denominator_relation}
\end{align}
where
\begin{equation}
 m_J^2
 =
 \frac{f_R-Rf_{RR}}{3f_{RR}}
 =
 \frac{f_R}{3f_{RR}}-\frac{R}{3}.
 \label{eq:quasistatic_mass_relation}
\end{equation}
Thus the standard observable denominator contains $k^2/a_J^2+m_J^2+R/3$, whereas the constant-background normal response contains $k^2/a_J^2+m_J^2$. In the slowly varying subhorizon regime, background-curvature terms such as $R/3$ are neglected relative to the physical wave number and scalaron scale. To that accuracy, the two denominator structures describe the same scalar response scale. The normal inverse itself is, in Fourier space,
\begin{equation}
 G_P(k,\widetilde a)
 \simeq
 -\frac{1}{3f_R}
 \frac{1}{k^2/\widetilde a^2+m_E^2}.
 \label{eq:Gp_quasistatic}
\end{equation}
Using $\widetilde a^2=f_Ra_J^2$ and $m_J^2=f_Rm_E^2$, the same operator may be written in Jordan variables as
\begin{equation}
 G_P(k,a_J)
 \simeq
 -\frac{1}{3}
 \frac{1}{k^2/a_J^2+m_J^2}.
 \label{eq:Gp_quasistatic_Jordan}
\end{equation}
Within the same quasistatic accuracy, the transition is controlled by the ratio of the physical wave number to the corresponding frame mass. In the heavy-scalar regime the response is suppressed; in the light-scalar regime it produces the familiar $1/3$ enhancement.

A direct intrinsic cross-check follows from the Jordan trace equation. On a slowly varying background, with $F\equiv f_R$ and $\delta F=f_{RR}\delta R$, its scalar part reduces to
\begin{equation}
 (\Box-m_J^2)\delta F\simeq\frac{\kapp^2}{3}\delta T,
 \label{eq:deltaF_scalar_eq}
\end{equation}
where
\begin{equation}
 m_J^2=\frac13\left(\frac{f_R}{f_{RR}}-R\right).
 \label{eq:Jordan_mass_deltaF}
\end{equation}
Since $\delta s=\delta F/F$, this is the same pole as in Eq.~\eqref{eq:L_quasistatic}, with $m_J^2=f_Rm_E^2$ and the corresponding conformal rescaling of the d'Alembertian. In projected variables the same mode is written as
\begin{equation}
 \delta s=-G_PB[\gamma]+\delta s_h,
 \label{eq:delta_s_Gp_matter}
\end{equation}
where the metric perturbation is itself sourced by $\delta T_{\mu\nu}$. Adding an independent matter term to the geometric tangent constraint would therefore double count the source.

The standard effective coupling and slip are therefore the observable closure of the same scalaron denominator. Their derivation still uses the complete metric--matter equations; the projection equation alone carries no separate matter source.

Diffeomorphism covariance is automatic. For an infinitesimal vector field,
\begin{equation}
 \gamma_{\mu\nu}=\Lie_\xi\gE_{\mu\nu},
 \qquad \sigma=\Lie_\xi s,
 \label{eq:gauge_lift}
\end{equation}
and
\begin{equation}
 B[\Lie_\xi\gE]+\cL_\Phi(\Lie_\xi s)=\Lie_\xi\hPf=0.
 \label{eq:gauge_projection_identity}
\end{equation}
Thus pure-gauge directions lie in the projected tangent space. A parent gauge functional pulls back as
\begin{equation}
 F_\mu^\star[\gE]=F_\mu[\gE,s_\star[\gE]],
 \label{eq:projected_gauge_functional}
\end{equation}
and can inherit $G_P$ only when it depends explicitly on the scalar.

A large scalaron mass makes the local Green response short ranged, one ingredient of chameleon screening. For extended bodies the exterior charge also depends on the nonlinear interior solution, matching, and possible thin-shell suppression; the normal operator diagnoses the local propagation scale rather than replacing that boundary-value problem.

\subsection{Reconstruction from expansion histories}

We finally return to the inverse problem at the background level. Reconstructing an $f(R)$ model from a prescribed expansion history is another manifestation of the same differential projection structure. The FLRW projection equation makes explicit the data that are hidden in the usual Jordan-frame reconstruction programme. In the Jordan frame one starts from a desired expansion history and solves for a function $f(R)$ that realizes it. In the projected Einstein description, the same construction becomes a problem of solving the nonlinear section equation for $s_\star[\gE]$. A section is not fixed until the Legendre branch, functional domain, boundary or Cauchy data, and global existence and uniqueness conditions are specified; the associated Green operator governs only the sourced linearized response around that section.

Let $N=\ln a_J$ be the Jordan-frame e-folding number and define
\begin{equation}
 G(N)=H_J(N)^2 .
 \label{eq:G_N_def}
\end{equation}
For a spatially flat background,
\begin{equation}
 R_J=6\left(H_JH_J'+2H_J^2\right)=3G'(N)+12G(N),
 \label{eq:R_N_G}
\end{equation}
where a prime denotes $\dd/\dd N$. The first modified Friedmann equation can be written as a second-order differential equation for $f(R)$ along the prescribed background. Using
\begin{equation}
 \frac{\dd f_R}{\dd t_J}=H_J\frac{\dd f_R}{\dd N}=H_J f_{RR}R_J'(N).
 \label{eq:fdot_N}
\end{equation}
one obtains
\begin{equation}
 0=-3G f_R+\frac12(R_Jf_R-f)-3G f_{RR}R_J'(N)+\kapp^2\rho_J(N).
 \label{eq:reconstruction_basic}
\end{equation}
Since
\begin{equation}
 R_J'(N)=3G''(N)+12G'(N),
 \label{eq:Rprime_N}
\end{equation}
Eq.~\eqref{eq:reconstruction_basic} gives a reconstruction equation once $N$ is expressed as a function of $R_J$ on a monotonic branch. Equivalently, using the standard form in terms of $H_J(N)$,
\begin{align}
 0&=-18\left(4H_J^3H_J'+H_J^2H_J'^2+H_J^3H_J''\right)f_{RR}
 \nonumber\\
 &\quad+3\left(H_J^2+H_JH_J'\right)f_R-\frac12f+\kapp^2\rho_J .
 \label{eq:reconstruction_standard}
\end{align}
This equation is intrinsic to the metric theory and has been widely used in cosmological reconstruction \cite{Nojiri:2006ri,Nojiri:2009kx,Nojiri:2010wj}. In the language of the present paper, it constructs a Jordan-frame metric-theory background directly on the projected field space.

The corresponding Einstein-side construction is more constrained than a generic scalar-tensor reconstruction. Given $\widetilde a(\widetilde t)$, the metric theory requires a scalar $s_\star$ satisfying Eq.~\eqref{eq:FLRW_projection_ODE}. Conversely, given a Jordan-frame solution $(a_J(t_J),R_J(t_J))$, one obtains
\begin{align}
 s&=\ln f_R(R_J),
 \\
 \widetilde a&=f_R(R_J)^{1/2}a_J,
 \qquad
 \dd\widetilde t=f_R(R_J)^{1/2}\dd t_J .
 \label{eq:Jordan_to_EF_background}
\end{align}

This is a forward map from a known Jordan solution. Conversely, prescribing $\widetilde a(\widetilde t)$ requires solving Eq.~\eqref{eq:FLRW_projection_ODE} and checking branch regularity, global existence, and the intended Jordan-equivalent data. The Jordan reconstruction equation and the Einstein-side section equation are complementary inverse problems: the former reconstructs $f(R)$ from a Jordan history, while the latter reconstructs or tests the scalar section for a fixed $f(R)$ and Einstein metric.

The distinction is relevant when an Einstein-frame expansion is prescribed externally, when only part of the parent equations is solved, or when an approximate trajectory is used. In those cases one must verify
\begin{equation}
 R_J[e^{-s}\gE]=X(e^s)
 \label{eq:cosmo_projection_condition_again}
\end{equation}
throughout the evolution. By contrast, every exact solution of the complete scalar--tensor parent derived from the same $f(R)$ satisfies this condition by the parent-shell identity. The normal operator is therefore a local test of the conditioning and invertibility of the chosen section, not a replacement for the full cosmological equations.

Near a spectral obstruction the metric-only Einstein-side reconstruction can become ill conditioned: a small change of the prescribed metric history may produce a large scalar response or fail the cokernel compatibility condition. This statement concerns the chosen coordinate graph. Stability of the physical cosmology still requires the coupled tensor, scalar, matter, and gauge perturbation system, together with its kinetic and gradient matrices and its initial data.

\section{Breakdown of the projected graph: kernel and cokernel obstructions}
\label{sec:caustics}

Nonlocal inversion is not the only problem produced by a curvature-dependent conformal map. The selected metric-only graph can also cease to exist when the normal linearization fails to be invertible on its chosen domain. This section separates that graph obstruction from Legendre degeneracy, physical instability, and zero modes of the action Hessian.

A metric-only Einstein section exists locally only when the linearization in the eliminated scalar direction is an isomorphism between the selected function spaces. In this context ``isomorphism'' means both existence and uniqueness: every admissible source must correspond to exactly one scalar response in the chosen domain. Failure can occur through noninjectivity, nonsurjectivity, or both. In a Fredholm realization these possibilities are measured by $\ker\cL_\Phi$ and $\ker\cL_\Phi^\dagger$, respectively. The implicit-function theorem then ceases to guarantee a unique smooth graph $s_\star[\widetilde g]$. We refer to this as a breakdown of the selected projected graph. The word ``caustic'' may be used only as an analogy after a nonlinear fold or bifurcation has been demonstrated; a zero of the linearization is merely a necessary diagnostic and is not, by itself, proof of multivaluedness.

On a constant projected background,
\begin{equation}
 \cL_\Phi
 =3\Phi(\BoxE-m_E^2),
 \qquad
 m_E^2=\frac{f_R-Rf_{RR}}{3f_Rf_{RR}}.
 \label{eq:constant_L_full}
\end{equation}
If the background is also on the constant parent scalar shell, $W_s=0$, then $W_{ss}=3m_E^2$. A vertical homogeneous solution obeys
\begin{equation}
 \cL_\Phi\omega=0,
 \qquad
 (\BoxE-m_E^2)\omega=0,
 \label{eq:vertical_zero_mode}
\end{equation}
with $m_E^2$ defined in Eq.~\eqref{eq:constant_L_full}. The sign of $m_E^2$ affects dynamical growth, but it does not change the principal symbol. On a globally hyperbolic Lorentzian spacetime, $\BoxE-m_E^2$ remains normally hyperbolic for either sign of the mass term, and the Cauchy problem is not rendered ill posed merely by $m_E^2<0$ \cite{Bar:2013bpw}. A negative mass squared can produce tachyonic growth; it is not automatically a failure of the retarded or advanced section.

The precise chart-failure condition is therefore
\begin{equation}
 \cL_\Phi:\mathcal D\longrightarrow\mathcal S
 \quad\hbox{is not invertible},
 \label{eq:spectral_caustic_full}
\end{equation}
for the stated domain $\mathcal D$, source space $\mathcal S$, and boundary or Cauchy prescription. The constant-mode condition
\begin{equation}
 m_E^2=0,
 \qquad
 f_R-Rf_{RR}=0,
 \qquad
 f_Rf_{RR}\neq0,
 \label{eq:homogeneous_fold_full}
\end{equation}
is only a homogeneous diagnostic. It produces an actual linearized obstruction only when the constant mode belongs to the chosen domain and satisfies its homogeneous boundary conditions.

For
\begin{equation}
 f(R)=R+\beta R^3,
 \qquad \beta>0,
 \label{eq:R_beta_R3}
\end{equation}
one has
\begin{equation}
 f_R=1+3\beta R^2,
 \qquad
 f_{RR}=6\beta R,
 \label{eq:R3_derivatives}
\end{equation}
and
\begin{equation}
 f_R-Rf_{RR}=1-3\beta R^2.
 \label{eq:R3_fold}
\end{equation}
Thus
\begin{equation}
 R_c=\pm\frac{1}{\sqrt{3\beta}}
 \label{eq:R3_Rc}
\end{equation}
are regular Legendre points with a massless constant vertical mode. They are candidate constant-mode degeneracies of boundary-value problems that admit that mode; they do not imply a universal graph breakdown for every Lorentzian or Dirichlet realization.

\subsection{De Sitter shells and dynamical stability}
\label{subsec:desitter_shells}

A vacuum constant-curvature solution satisfies
\begin{equation}
 R_0f_R(R_0)-2f(R_0)=0,
 \label{eq:deSitter_condition}
\end{equation}
with
\begin{equation}
 s_0=\ln f_R(R_0),
 \qquad
 \widetilde R_0=\frac{R_0}{f_R(R_0)}.
 \label{eq:s0_dS}
\end{equation}
The normal operator is
\begin{equation}
 \cL_{\Phi,0}
 =3f_R(R_0)(\BoxE-m_E^2),
 \label{eq:L_dS}
\end{equation}
where
\begin{equation}
 m_E^2
 =\frac{f_R(R_0)-R_0f_{RR}(R_0)}
 {3f_R(R_0)f_{RR}(R_0)},
 \qquad
 m_J^2=f_R(R_0)m_E^2.
 \label{eq:mass_dS}
\end{equation}
On the usual viable branch $f_R>0$, $f_{RR}>0$, positivity of the numerator gives the standard absence of a tachyonic scalaron on de Sitter \cite{Faraoni:2007yn,DeFelice:2010aj}. This is a physical stability condition. Local invertibility of the projected section is a separate operator-domain question. In particular, $m_E^2=0$ obstructs a section only if the corresponding constant mode is admissible.

For the Starobinsky model
\begin{equation}
 f(R)=R+\frac{R^2}{6M^2},
 \label{eq:Starobinsky_M}
\end{equation}
\begin{equation}
 f_R=1+\frac{R}{3M^2},
 \qquad
 f_{RR}=\frac{1}{3M^2},
 \label{eq:Starobinsky_derivs}
\end{equation}
and the parent potential is
\begin{equation}
 W(s)=\frac{3M^2}{2}(1-\ee^{-s})^2.
 \label{eq:Starobinsky_W}
\end{equation}
The canonical scalar
 $\varphi=\sqrt{3/2}\,s/\kapp$ has
\begin{equation}
 V(\varphi)=\frac{3M^2}{4\kapp^2}\left(1-\ee^{-\sqrt{2/3}\kapp\varphi}\right)^2.
 \label{eq:Starobinsky_V}
\end{equation}
The scalaron remains the physical spin-zero metric mode. In the parent it is represented by an independent coordinate; in the metric theory its fluctuation is represented by the projected tangent data. Standard on-shell inflationary observables are unchanged under a consistent mapping of variables and states, whereas off-shell Gaussian calculations must use the projected domain and complete pulled-back Hessian.

\subsection{Euclidean spectra and failure of the graph section}
\label{sec:euclidean_spectra}

On a compact Euclidean constant-curvature background, let
\begin{equation}
 -\BoxE Y_n=\lambda_nY_n,
 \qquad \lambda_n\ge0,
 \label{eq:laplace_spectrum}
\end{equation}
then
\begin{equation}
 \cL_\Phi Y_n
 =-3\Phi(\lambda_n+m_E^2)Y_n,
 \label{eq:L_spectrum_euclidean}
\end{equation}
and the chosen elliptic realization has a zero mode when
\begin{equation}
 \lambda_n+m_E^2=0.
 \label{eq:zero_mode_euclidean}
\end{equation}
On a four-sphere of radius $a$ one has
\begin{equation}
 \lambda_\ell=\frac{\ell(\ell+3)}{a^2},
 \qquad \ell=0,1,2,\ldots,
 \label{eq:S4_eigenvalues}
\end{equation}
so a normal zero mode occurs if
\begin{equation}
 m_E^2=-\frac{\ell(\ell+3)}{a^2}.
 \label{eq:S4_zero_condition}
\end{equation}
This is an obstruction to the unique graph section and to the ordinary normal Jacobian. It is not, by itself, a zero eigenvalue of the projected action Hessian.

Let $\{\psi_A\}$ span $\ker\cL_\Phi$, let $G_P$ be a generalized inverse on a complement, and suppose the compatibility condition \eqref{eq:tangent_fredholm_condition} holds. The linearized constrained fluctuation then has the form
\begin{equation}
 \sigma=-G_PB[\gamma]+\sum_Ac^A\psi_A.
 \label{eq:zero_mode_tangent_decomposition}
\end{equation}
The Gaussian integral is correspondingly of the schematic form
\begin{align}
Z_{\rm proj}^{(2)}\propto{}&
\int\mathcal D\gamma\,\prod_A\dd c^A\,
\delta[\Pi_{\rm coker}B\gamma]
\nonumber\\
&\times
\exp\!\left\{
\frac{i}{2}
H_{\rm constr}^{(2)}
\left[(\gamma,-G_PB\gamma+c^A\psi_A)^2\right]
\right\},
\label{eq:projected_gaussian_zero_modes}
\end{align}
with the appropriate measure, ghost, and normal Jacobian factors understood. The coefficients $c^A$ are integrated with the quadratic form induced by the action. They become collective coordinates requiring primed determinants only if that quadratic form also has genuine zero directions, for example because of a symmetry. A prime on a determinant means that exact zero eigenvalues are omitted from the determinant and treated separately. A collective coordinate is a parameter used to integrate explicitly over such an exact flat direction, typically when it is generated by a symmetry. A normal zero mode alone does not imply a flat direction of the projected Hessian.

Away from these operator-domain obstructions, a regular variational section can be substituted into the parent action. We now derive its metric-only Einstein-side quadratic kernel.

\section{Metric-only Einstein-side action and nonlocal form factors}
\label{sec:nonlocal_action}

On a regular branch and for a specified variational boundary problem, the pulled-back action
\begin{equation}
 S_\star[\gE]
 =S_E[\gE,s_\star[\gE]]
 \label{eq:Sstar_nonlocal}
\end{equation}
is the exact pullback of the parent action to the graph sector selected by that branch, functional domain, and boundary prescription. It represents the corresponding sector of the intrinsic metric theory, but it is not a single global metric-only parametrization of the complete scalaron phase space when distinct admissible homogeneous data lie above the same Einstein metric. A global graph can also fail because the forward map has multiple off-shell preimages or because the normal graph construction becomes singular. The reduced action is generically nonlocal because $s_\star[\gE]$ is defined by a differential inverse problem.

Around a regular background,
\begin{equation}
 \gE_{\mu\nu}=\bar\gE_{\mu\nu}+\gamma_{\mu\nu},
 \quad
 s_\star=\bar s-G_P^{\rm var}B[\gamma]+\cO(\gamma^2),
 \label{eq:sstar_expansion}
\end{equation}
where $G_P^{\rm var}$ denotes the inverse associated with the variational problem. The quadratic action is
\begin{align}
S_\star^{(2)}[\gamma]
={}&\frac12\delta^2S_E
[(\gamma,-G_P^{\rm var}B\gamma)^2]
\nonumber\\
&-\frac12
\left\langle E_s,
G_P^{\rm var}\delta^2\hPf
[(\gamma,-G_P^{\rm var}B\gamma)^2]
\right\rangle .
\label{eq:Sstar_quadratic_nonlocal}
\end{align}
This is the complete quadratic pullback on the selected branch. On the parent shell the second line vanishes, but the first line still contains the nonlocal tangent lift.

For the flat projected $R+\alpha R^2$ background,
\begin{equation}
 \widetilde R^{(1)}+3(\BoxE-m_s^2)s=0,
\end{equation}
and the scalar quadratic action is
\begin{equation}
 S_s^{(2)}
 =\frac{3}{4\kapp^2}
 \int\dd^4x\,
 s(\BoxE-m_s^2)s .
 \label{eq:flat_scalar_parent_quadratic}
\end{equation}
For a self-adjoint variational realization of the inverse, exact elimination gives
\begin{equation}
 S_{\star,0}^{(2)}
 =S_{\rm EH}^{(2)}[\gamma]
 +\frac{1}{12\kapp^2}
 \int\dd^4x\,
 \widetilde R^{(1)}
 \frac{1}{\BoxE-m_s^2}
 \widetilde R^{(1)} .
 \label{eq:nonlocal_scalar_curvature_kernel}
\end{equation}
The linearized scalar curvature is gauge invariant on the flat background, and Eq.~\eqref{eq:nonlocal_scalar_curvature_kernel} contains the exact scalaron denominator in the nonlocal spin-zero form factor. This denominator should not, by itself, be identified with a pole of the vacuum Einstein-metric propagator: the pole content must be read from the complete constrained quadratic system and from consistently pulled-back Jordan-frame or matter observables. The Einstein--Hilbert sector retains the massless spin-two pole. Equation~\eqref{eq:nonlocal_scalar_curvature_kernel} is the exact Gaussian pullback on the selected graph sector, not an additional physical theory.

The same quadratic kernel follows directly from the constrained Gaussian integral. Define
\begin{equation}
 D_s\equiv\BoxE-m_s^2
 \label{eq:Ds_definition}
\end{equation}
and consider an invertible self-adjoint realization of $D_s$. Holding the metric measure, gauge factors, and pushforward Jacobian fixed, and using the factorized translation-invariant scalar measure displayed here, one obtains
\begin{align}
&\int\mathcal Ds\,
 \delta\!\left[
  \widetilde R^{(1)}+3D_ss
 \right]
 \exp\!\left\{
  \frac{3i}{2}\langle s,D_ss\rangle
 \right\}
\nonumber\\
&\qquad =
  \bigl|\det(3D_s)\bigr|^{-1}
 \exp\!\left\{
  \frac{i}{6}
  \left\langle
   \widetilde R^{(1)},
   D_s^{-1}\widetilde R^{(1)}
  \right\rangle
 \right\} .
\label{eq:explicit_constrained_Gaussian_reduction}
\end{align}
For the real delta functional used here, the absolute value follows from
\begin{equation}
 \delta[A\phi]=|\det A|^{-1}\delta[\phi].
 \label{eq:delta_functional_determinant_identity}
\end{equation}
If an oriented determinant or a determinant with a separately prescribed phase is used instead, the corresponding contour or spectral convention must be specified. The same convention applies to the primed determinant appearing below when zero modes are present.

Because the pairing in Eq.~\eqref{eq:scalar_pairing_def} already contains $1/(2\kapp^2)$, the final exponent is explicitly
\begin{equation}
 \frac{i}{12\kapp^2}\int\dd^4x\,\widetilde R^{(1)}D_s^{-1}\widetilde R^{(1)} .
 \label{eq:explicit_Gaussian_exponent_integral}
\end{equation}
Thus the nonlocal kernel in Eq.~\eqref{eq:nonlocal_scalar_curvature_kernel} and a normal determinant are produced simultaneously by this constrained scalar integration. The specific determinant in Eq.~\eqref{eq:explicit_constrained_Gaussian_reduction} is the result for the displayed factorized scalar measure. In the metric theory, the pushforward of the Jordan metric measure may supply normal factors that modify or compensate it; a complete result also requires the projected gauge and ghost sectors, regularization, and the zero-mode prescription. If $D_s$ has a kernel, let $D_s^+$ denote the generalized inverse on a chosen complementary subspace. In this flat constant-parent-shell example, $\mathcal L_\Phi=3D_s$ is proportional to the scalar vertical Hessian, so $D_s\psi_A=0$ implies $S_s^{(2)}[\psi_A]=0$. The scalar integration is then schematically
\begin{align}
Z_s^{(2)}\propto{}&
  \bigl|\det{}'(3D_s)\bigr|^{-1}\delta\!\left[\Pi_{\ker D_s}\widetilde R^{(1)}\right]
 \operatorname{Vol}(\ker D_s)
\nonumber\\
&\times
 \exp\!\left\{
  \frac{i}{12\kapp^2}
  \int\dd^4x\,
  \widetilde R^{(1)}D_s^+
  \widetilde R^{(1)}
 \right\} ,
\label{eq:Gaussian_zero_mode_compatibility}
\end{align}
again up to the remaining metric, gauge, and measure factors. The prime acts on the invertible complement. Hence the zero mode first enforces a compatibility condition on the metric source, while its scalar kernel coordinate is a flat Gaussian direction and contributes a formal kernel volume. Higher-order terms, additional section data, symmetry quotients, or collective-coordinate methods are required to treat that direction beyond the quadratic approximation. This special flatness does not contradict the general statement around Eq.~\eqref{eq:projected_gaussian_zero_modes}, where a normal zero mode need not coincide with a zero mode of the complete projected action Hessian.

\subsection{Solution operators, variational kernels, and causality}
\label{sec:EFT_comments}

The word ``Green function'' refers to different objects in different problems. The decaying Yukawa inverse in Sec.~\ref{sec:weakfield} is an elliptic static Green function. The retarded operator in cosmology is a causal solution map. A Euclidean or self-adjoint inverse defines an ordinary variational quadratic form, while the Feynman inverse appears in an in--out effective action.

A purely retarded kernel is not symmetric,
\begin{equation}
 (G_{\rm ret})^\dagger=G_{\rm adv},
\end{equation}
and cannot simply be inserted as the kernel of a standard single-history action whose second variation is symmetric. Causal expectation-value equations are obtained with an in--in or Schwinger--Keldysh prescription, or by the corresponding causal continuation of a Euclidean effective action \cite{Barvinsky:2014lja}. Thus causal response and variational nonlocality are compatible, but their inverse prescriptions must be kept distinct.

\subsection{Derivative expansion and effective-field-theory interpretation}

Below the scalaron scale,
\begin{equation}
 (1-6\alpha\BoxE)^{-1}
 =1+6\alpha\BoxE+(6\alpha\BoxE)^2+\cdots
 \label{eq:EFT_form_factor}
\end{equation}
provides a controlled derivative expansion. Applied to Eq.~\eqref{eq:nonlocal_scalar_curvature_kernel}, it generates local operators in the scalar-curvature sector, schematically $\widetilde R^2$, $\widetilde R\BoxE\widetilde R$, and higher derivatives. Eliminating the spin-zero scalaron does not generate a Weyl-squared term in the quadratic spin-two sector.

A finite truncation used perturbatively within $|\BoxE|\ll m_s^2$ is an effective theory and does not add new physical initial data order by order. If the truncated higher-derivative polynomial is instead promoted to an exact nonperturbative kinetic operator, additional roots can appear and may be ghost-like, tachyonic, or complex. Such roots usually lie at or above the scale where the derivative expansion has lost validity and should not be interpreted as predictions of the controlled EFT \cite{Simon:1990ic,Burgess:2003jk, Solomon:2017nlh}. The unexpanded kernel retains the exact scalaron denominator in the nonlocal form factor and avoids the additional polynomial roots generated when a finite derivative truncation is promoted to an exact kinetic operator. The physical pole interpretation must be made at the level of the complete constrained system or of consistently transformed observables.

At the quantum level, the constrained parent integral and the metric-only nonlocal representation can describe the same theory only when the measure, normal Jacobian, gauge fixing, boundary or state prescription, and regularization are transformed consistently. Enforcing the auxiliary constraint is essential, but the present construction does not by itself prove all-loop renormalizability or unitarity. The reliable conclusion is more specific: the projected and unrestricted parent Gaussian theories have different fluctuation domains and quadratic kernels off shell, while a consistent constrained formulation preserves the physical scalaron pole and the common on-shell observables.

\section{Beyond metric \texorpdfstring{$f(R)$}{f(R)}: extensions and outlook}
\label{sec:beyond_metric_fr}

The preceding theorem applies directly to the nondegenerate metric Ricci-scalar class. Its broader lesson is an audit procedure rather than a universal claim: identify the independent carrier of the conformal factor, derive the constraints produced when that carrier is eliminated, and test the complete response $-\mathcal N^{-1}B$. A forward image of a known solution does not by itself establish a closed local variational theory in the reduced variables.

\subsection{Palatini and metric-affine theories}

In Palatini $f(\mathcal R)$ gravity the metric and connection are independent. The trace relation
\begin{equation}
 f_{\mathcal R}\mathcal R-2f=\kapp^2T
 \label{eq:Palatini_trace_relation}
\end{equation}
is algebraic on a regular branch \cite{Olmo:2011uz}. With $F=f_{\mathcal R}$ and $h_{\mu\nu}=Fg_{\mu\nu}$, the connection is Levi--Civita for $h_{\mu\nu}$, and
\begin{equation}
 G_{\mu\nu}(h)=\frac{\kapp^2}{F}T_{\mu\nu}
 -\frac{F\mathcal R-f}{2F^2}h_{\mu\nu}.
 \label{eq:Palatini_Einstein_h}
\end{equation}
The auxiliary metric therefore obeys Einstein-like equations with both a rescaled matter source and a matter-dependent potential term. In the scalar--tensor description, pure Palatini $f(\mathcal R)$ corresponds to $\omega_{\rm BD}=-3/2$, for which the scalar wave operator degenerates and
\begin{equation}
 2V(\phi)-\phi V_\phi(\phi)=\kapp^2T
 \label{eq:Palatini_scalar_algebraic_constraint}
\end{equation}
is algebraic. Thus its scalar normal problem is local before the connection is eliminated. Derivatives of matter can nevertheless enter after $F(T)$ and the connection are substituted into a metric-only representation.

In general metric-affine theories the connection can carry torsion or nonmetricity \cite{Hehl:1994ue}. A conformal transformation may be local on $(g_{\mu\nu},\Gamma^\lambda{}_{\mu\nu})$ but become nonlocal after $\Gamma^\lambda{}_{\mu\nu}$ is projected out. Whether that reduction is algebraic or differential depends on the action and on whether the connection is auxiliary or propagating. The scalar normal operator derived here is therefore a template for the required coupled calculation, not a result for the whole metric-affine class.

\subsection{Comparison of conformal constructions}
\label{subsec:forward_images_vs_frames}

The inverse problem is therefore theory dependent. In metric $f(R)$ the carrier is the curvature of the same metric being reconstructed, and the complete response is the nonpolynomial scalar operator derived above. In pure Palatini $f(\mathcal R)$, the scalar carrier is fixed algebraically by matter before the connection is removed. Hybrid and general metric-affine models require a coupled scalar--connection analysis. An independent scalar--tensor conformal factor, by contrast, remains algebraically invertible on the enlarged field space.

This comparison separates two operations that are often conflated. Applying a conformal rule to a known solution produces a forward image. Constructing a reduced off-shell theory requires more: one must identify the constraint set, its tangent space, and the inverse of the eliminated-direction linearization. Only the latter question determines whether the transformed variables form a local unconstrained chart.

For orientation, the main cases may be summarized as follows:
\begin{center}
\renewcommand{\arraystretch}{1.15}
\begin{tabular}{p{0.27\columnwidth}p{0.65\columnwidth}}
\hline
\textbf{Construction} & \textbf{Reduced inverse problem} \\
\hline
Metric $f(R)$ & Differential scalar constraint; the complete response is
generically nonlocal and is not a finite-order differential operator.\\
Independent scalar--tensor & The scalar is retained; the Weyl inverse is algebraic on a regular branch. \\
Palatini $f(\mathcal R)$ & Algebraic scalar constraint before connection elimination; matter derivatives may appear afterwards. \\
Hybrid/metric-affine & Coupled scalar--connection constraints; locality is
determined by a normal-operator matrix. \\
\hline
\end{tabular}
\end{center}
Here a \emph{normal-operator matrix} is simply the coupled linear system in all eliminated scalar, vector, tensor, or connection directions. Its inverse must be applied to the sources generated by variations of the retained fields; testing only one diagonal block is insufficient.

\subsection{Hybrid and higher-curvature constructions}

Hybrid metric--Palatini gravity contains both curvatures
\cite{Harko:2011nh}, for example
\begin{equation}
 S=\frac{1}{2\kapp^2}\int\dd^4x\sqrt{-g}\,[R+f(\mathcal R)]
 +S_m[g,\psi].
 \label{eq:hybrid_action}
\end{equation}
Its scalar--tensor representation contains a propagating scalar even though the pure Palatini scalar is algebraic. A metric-only Einstein-side reduction would require the coupled scalar--connection projection, the corresponding normal-operator matrix, its kernel and cokernel, and the pushforward measure. That construction is model dependent and is left for future work.

For conformal factors depending on several curvature invariants,
\begin{equation}
 \gE_{\mu\nu}=F(R,R_{\alpha\beta}R^{\alpha\beta},
 R_{\alpha\beta\rho\sigma}R^{\alpha\beta\rho\sigma},\ldots)g_{\mu\nu},
 \label{eq:higher_curvature_factor}
\end{equation}
auxiliary localization generally produces a matrix of scalar and tensor constraints. The locality question is then controlled by the functional rank and symbol of the full matrix response, not by the differential order of one block. Kernel or cokernel modes signal failure of the selected linear graph; a nonlinear fold still requires a separate bifurcation analysis.

An $R^2$ sector can be localized by a scalar, whereas a Weyl-squared sector carries a massive spin-two structure and requires a tensor auxiliary field or a tensor block in the normal matrix \cite{Stelle:1977ry,Hindawi:1995an, Rodrigues:2011zi}. Curvature--matter conformal factors may instead remain local when the carrier is algebraically fixed by matter. The same logic applies to curvature--matter couplings. If the conformal factor is algebraically determined by matter, the reduced section may remain local. If it depends on curvature composites of the metric being reconstructed, a differential projection reappears and the complete matrix response must be examined. A kernel or cokernel then signals failure of the selected linear section, but not automatically a nonlinear bifurcation or a physical instability.

The general prescription is therefore to retain independent carriers until all projection constraints and their complete composite response have been computed. The scalar construction developed here applies directly to the trace sector; extensions with tensor auxiliaries require the corresponding matrix normal problem. Exceptional cancellations must be demonstrated rather than inferred from the formal derivative order.

\section{Discussion and conclusions}
\label{sec:discussion}

We have established a precise locality obstruction for conformal transformations whose conformal factor depends nontrivially on the Ricci scalar of the metric being transformed. Although such transformations are local in the forward direction for a known metric, their inversion requires the reconstruction of the curvature of the unknown preimage metric. The inverse problem is therefore intrinsically differential.
 
By introducing an independent auxiliary scalar, we localized the transformation on an enlarged field space and derived the complete inverse metric response. Its nonpolynomial symbol shows that the original metric cannot, in general, be recovered from the transformed metric through a finite-order local differential relation. Thus, on a nondegenerate branch, there is no generic differentiable finite-jet metric-only inverse on an open set of unrestricted configurations.

This result concerns locality rather than absolute invertibility. Branchwise functional inverses may exist after specifying a functional domain, boundary or Cauchy data, and a prescription for homogeneous modes. A valid transformation between known solutions should therefore not be confused with a local change of coordinates on the full off-shell metric configuration pace.
 
Metric $f(R)$ gravity provides the principal physical realization of this distinction. Its scalar--tensor representation is a local parent theory in which the conformal factor is carried by an independent scalar. The original metric theory occupies only a differential constraint set inside this enlarged field space. A metric-only Einstein-side formulation consequently requires solving the constraint that reconstructs the scalar from the transformed metric.
 
The standard classical equivalence is unaffected. Exact regular solutions of the parent theory satisfy the projection automatically, and the scalaron remains the physical spin-zero degree of freedom encoded in the higher-derivative metric sector. The distinction arises off shell, where arbitrary parent configurations, variational deformations, Gaussian fluctuations, and functional-integral histories need not correspond to configurations of the original metric theory.
 
The normal operator provides the linearized description of this projection. Its Green inverse determines the sourced scalar response, while homogeneous solutions encode complementary boundary or initial data. On common physical backgrounds, its pole is the scalaron pole and it directly governs the weak-field Yukawa response. In the quasistatic cosmological regime, the same scalar response scale organizes the scale-dependent observables after the background-curvature terms neglected in that approximation have been consistently removed.
 
At the variational level, the metric-theory Hessian is obtained by pulling back the parent action to the differential section. In addition to the parent Hessian restricted to allowed tangent fluctuations, it contains an off-shell embedding contribution that vanishes on the common classical shell. The quadratic model further shows how constrained Gaussian elimination generates the corresponding nonlocal scalaron kernel, normal determinant for the chosen measure, and zero-mode compatibility condition. A complete one-loop comparison nevertheless requires the induced functional measure, constraint Jacobian, gauge and ghost sectors, regularization, and state or boundary prescription.
 
Several directions for future research follow naturally. A first priority is the construction of fully gauge-fixed projected one-loop operators, including the transformed measure and the associated ghost and Jacobian contributions. This would allow a direct quantum comparison between the constrained metric theory and its unrestricted scalar--tensor parent.
 
A second direction is the global nonlinear projection problem. It remains to determine when a branchwise inverse is unique, when distinct off-shell preimages coexist, and when the metric-only graph develops genuine folds or bifurcations. These questions will require analytical and numerical methods beyond the linearized normal operator.
 
The framework should also be extended systematically to Palatini, hybrid metric--Palatini, metric-affine, and more general higher-curvature theories. In these cases the eliminated variables may include scalar, connection, or tensorial degrees of freedom, and the corresponding inverse problem will generally be governed by coupled normal-operator matrices.
 
Finally, it would be valuable to identify exceptional symmetry-reduced sectors or special conformal factors for which a local inverse response may survive. Such cases would not invalidate the generic obstruction, but could provide exactly solvable examples and clarify the boundary between local and intrinsically nonlocal field redefinitions.

The broader lesson is that locality on an enlarged parent field space need not survive elimination of the field carrying the conformal factor. Accordingly, a forward map between known configurations, a local transformation of a parent theory, and a local change of variables on the original off-shell configuration space are distinct constructions. Curvature-dependent conformal transformations should therefore not be used interchangeably in these three senses.

For the nondegenerate class studied here, treating the curvature-dependent rule naively as an ordinary metric field redefinition can change the variational and fluctuation problem: the transformed variations are restricted by a differential projection, its inverse depends on functional and boundary or Cauchy data, and the metric-theory Hessian is the pullback to that projected domain rather than the unrestricted parent Hessian. The same distinction propagates to effective actions, semiclassical expansions, and functional integrals. More generally, curvature-dependent transformations in modified gravity should therefore be audited by examining their complete inverse response before they are interpreted as genuine changes of off-shell field variables.

\begin{acknowledgments}
The authors acknowledge support from FCT through the research grant UID/04434/2025. FSNL also acknowledges support from the FCT Scientific Employment Stimulus contract with reference CEECINST/00032/2018.
\end{acknowledgments}

\appendix

\section{Derivation of the normal operator}
\label{app:normal_operator}

The projection is
\begin{equation}
 \hPf=\Phi Q-X(\Phi),
 \qquad
 \Phi=\ee^s,
 \label{eq:app_Pf_PhiQ}
\end{equation}
where
\begin{equation}
 Q=\RE+3\BoxE s-\frac32(\nablaE s)^2 .
 \label{eq:app_Q_def}
\end{equation}
The quantity $Q$ is precisely the combination that appears in the conformal transformation of the Ricci scalar, $R[g] = \Phi Q$, so that the projection $\hPf=0$ is simply the auxiliary condition $R[g]=X(\Phi)$ expressed in Einstein-frame variables.

At fixed $\gE$,
\begin{equation}
 \delta_s Q=3\BoxE\sigma-3\nablaE^\mu s\nablaE_\mu\sigma .
 \label{eq:app_delta_Q}
\end{equation}
Also $\delta\Phi=\Phi\sigma$. Hence
\begin{align}
 \delta_s\hPf
 &=\Phi\sigma Q+\Phi\delta_sQ-X_\Phi\Phi\sigma
 \\
 &=\Phi\left[Q-X_\Phi\right]\sigma
 +3\Phi\left(\BoxE-\nablaE^\mu s\nablaE_\mu\right)\sigma .
 \label{eq:app_delta_Pf_before_shell}
\end{align}
On the projected section $\Phi Q=X$, so $Q=X/\Phi$. Therefore
\begin{equation}
 \delta_s\hPf=\left[3\Phi(\BoxE-\nablaE^\mu s\nablaE_\mu)+X-\Phi X_\Phi\right]\sigma,
 \label{eq:app_Lphi_result}
\end{equation}
which is Eq.~\eqref{eq:Lphi_full}.

The result shows that the normal operator consists of a second-order differential part, proportional to $\BoxE$, and the algebraic term $X-\Phi X_\Phi$. The relation of this term to the Einstein-frame potential must be stated with care. From
\begin{align}
W_s&=\frac{X}{\Phi}-2W,\\
W_{ss}&=X_\Phi-\frac{3X}{\Phi}+\frac{4U}{\Phi^2},
\label{eq:Wss_general_identity}
\end{align}
one sees that $X-\Phi X_\Phi=-\Phi W_{ss}$ is not an off-shell identity on a generic constant projected background. It holds when the background also satisfies the constant parent scalar equation $W_s=0$, for which $W_{ss}=X_\Phi-X/\Phi$. On that common shell the normal operator is proportional to the scalar vertical Hessian and its pole is the scalaron pole. Away from the scalar shell, $X-\Phi X_\Phi$ remains the correct projection coefficient, while the off-shell parent scalar Hessian contains the additional terms displayed above.

\section{Second variation and embedding term}
\label{app:second_variation}

Let $\iota:\gE\mapsto(\gE,s_\star[\gE])$ be the section. In condensed notation,
\begin{equation}
 S_\star=S_E\circ\iota .
 \label{eq:app_Sstar_iota}
\end{equation}
The first variation is
\begin{equation}
 \delta S_\star[\gamma]
 =\delta S_E[\eta],
 \qquad
 \eta=(\gamma,\delta s_\star[\gamma]).
 \label{eq:app_first_var}
\end{equation}
The second variation contains the variation of the tangent map:
\begin{equation}
 \delta^2S_\star[\gamma_1,\gamma_2]
 =\delta^2S_E[\eta_1,\eta_2]
 +\left\langle E_s,\delta^2s_\star[\gamma_1,\gamma_2]\right\rangle.
 \label{eq:app_second_var}
\end{equation}
The second derivative of the section is determined by differentiating $\hPf\circ\iota=0$:
\begin{equation}
 \cL_\Phi\delta^2s_\star+
 \delta^2\hPf[\eta_1,\eta_2]=0.
 \label{eq:app_second_proj}
\end{equation}
Thus, on a regular section,
\begin{equation}
 \delta^2s_\star=-G_P\delta^2\hPf[\eta_1,\eta_2],
 \label{eq:app_second_s}
\end{equation}
which gives Eq.~\eqref{eq:Hsplit_full}. At a kernel or cokernel obstruction, this equation must instead be projected to a complementary subspace and supplemented by the compatibility condition \eqref{eq:tangent_fredholm_condition}.

This derivation shows that the embedding correction is a direct consequence of the constraint $\hPf=0$. The second derivative of the section is forced by the second variation of the projection, and the Green operator $G_P$ propagates this information into the pulled-back Hessian. Geometrically, the term $\langle E_s,\delta^2 s_\star\rangle$ is the inner product of the normal gradient of the action with the second fundamental form of the constraint surface $\mathcal{M}_f$. In finite dimensions, it is the standard correction to the Hessian when a function on a curved constraint set is expressed in ambient coordinates. That this term survives off shell but vanishes on the parent equations of motion is the origin of the on-shell agreement and off-shell mismatch between the metric and parent Hessians.

\bibliographystyle{apsrev4-2}
\bibliography{biblio}

@article{Chisholm:1961tha,
  author  = {Chisholm, J. S. R.},
  title   = {Change of Variables in Quantum Field Theories},
  journal = {Nucl. Phys.},
  volume  = {26},
  pages   = {469--479},
  year    = {1961},
  doi     = {10.1016/0029-5582(61)90106-7}
}

@article{Kamefuchi:1961sb,
  author  = {Kamefuchi, S. and O'Raifeartaigh, L. and Salam, Abdus},
  title   = {Change of Variables and Equivalence Theorems in Quantum Field Theories},
  journal = {Nucl. Phys.},
  volume  = {28},
  pages   = {529--549},
  year    = {1961},
  doi     = {10.1016/0029-5582(61)90056-6}
}

@article{Dicke:1961gz,
  author  = {Dicke, Robert H.},
  title   = {Mach's Principle and Invariance under Transformation of Units},
  journal = {Phys. Rev.},
  volume  = {125},
  pages   = {2163--2167},
  year    = {1962},
  doi     = {10.1103/PhysRev.125.2163}
}

@article{Falls:2018olk,
    author = "Falls, Kevin and Herrero-Valea, Mario",
    title = "{Frame (In)equivalence in Quantum Field Theory and Cosmology}",
    eprint = "1812.08187",
    archivePrefix = "arXiv",
    primaryClass = "hep-th",
    doi = "10.1140/epjc/s10052-019-7070-3",
    journal = "Eur. Phys. J. C",
    volume = "79",
    number = "7",
    pages = "595",
    year = "2019"
}

@article{BandaraGoffengSaratchandran2023,
    author = "Bandara, Lashi and Goffeng, Magnus and Saratchandran, Hemanth",
    title = "{Realisations of elliptic operators on compact manifolds with boundary}",
    eprint = "2104.01919",
    archivePrefix = "arXiv",
    primaryClass = "math.DG",
    journal = "Adv. Math.",
    volume = "420",
    pages = "108968",
    year = "2023",
    doi = "10.1016/j.aim.2023.108968"
}

@article{Faraoni:1998qx,
  author        = {Faraoni, Valerio and Gunzig, Edgard and Nardone, Pasquale},
  title         = {Conformal Transformations in Classical Gravitational Theories and in Cosmology},
  journal       = {Fund. Cosmic Phys.},
  volume        = {20},
  pages         = {121--175},
  year          = {1999},
  eprint        = {gr-qc/9811047},
  archivePrefix = {arXiv}
}

@book{FujiiMaedaBook,
  author    = {Fujii, Yasunori and Maeda, Kei-ichi},
  title     = {The Scalar-Tensor Theory of Gravitation},
  publisher = {Cambridge University Press},
  address   = {Cambridge},
  year      = {2003},
  doi       = {10.1017/CBO9780511535093},
  isbn      = {978-0-521-81159-0}
}

@article{Stelle:1977ry,
  author  = {Stelle, K. S.},
  title   = {Renormalization of Higher-Derivative Quantum Gravity},
  journal = {Phys. Rev. D},
  volume  = {16},
  pages   = {953--969},
  year    = {1977},
  doi     = {10.1103/PhysRevD.16.953}
}

@article{Starobinsky:1980te,
  author  = {Starobinsky, Alexei A.},
  title   = {A New Type of Isotropic Cosmological Models without Singularity},
  journal = {Phys. Lett. B},
  volume  = {91},
  pages   = {99--102},
  year    = {1980},
  doi     = {10.1016/0370-2693(80)90670-X}
}

@article{Whitt:1984pd,
  author  = {Whitt, Brian},
  title   = {Fourth-Order Gravity as General Relativity plus Matter},
  journal = {Phys. Lett. B},
  volume  = {145},
  pages   = {176--178},
  year    = {1984},
  doi     = {10.1016/0370-2693(84)90332-0}
}

@article{Maeda:1988ab,
  author  = {Maeda, Kei-ichi},
  title   = {Towards the Einstein-Hilbert Action via Conformal Transformation},
  journal = {Phys. Rev. D},
  volume  = {39},
  pages   = {3159--3162},
  year    = {1989},
  doi     = {10.1103/PhysRevD.39.3159}
}

@article{Teyssandier:1983zz,
  author  = {Teyssandier, Pierre and Tourrenc, Philippe},
  title   = {The Cauchy Problem for the $R+R^2$ Theories of Gravity without Torsion},
  journal = {J. Math. Phys.},
  volume  = {24},
  pages   = {2793--2799},
  year    = {1983},
  doi     = {10.1063/1.525659}
}

@article{Wands:1993uu,
  author        = {Wands, David},
  title         = {Extended Gravity Theories and the Einstein-Hilbert Action},
  journal       = {Class. Quant. Grav.},
  volume        = {11},
  pages         = {269--280},
  year          = {1994},
  doi           = {10.1088/0264-9381/11/1/025},
  eprint        = {gr-qc/9307034},
  archivePrefix = {arXiv}
}

@article{Magnano:1993bd,
  author        = {Magnano, Guido and Sokolowski, Leszek M.},
  title         = {Physical Equivalence between Nonlinear Gravity Theories and a General-Relativistic Self-Gravitating Scalar Field},
  journal       = {Phys. Rev. D},
  volume        = {50},
  pages         = {5039--5059},
  year          = {1994},
  doi           = {10.1103/PhysRevD.50.5039},
  eprint        = {gr-qc/9312008},
  archivePrefix = {arXiv}
}

@article{Sotiriou:2008rp,
  author        = {Sotiriou, Thomas P. and Faraoni, Valerio},
  title         = {$f(R)$ Theories of Gravity},
  journal       = {Rev. Mod. Phys.},
  volume        = {82},
  pages         = {451--497},
  year          = {2010},
  doi           = {10.1103/RevModPhys.82.451},
  eprint        = {0805.1726},
  archivePrefix = {arXiv},
  primaryClass  = {gr-qc}
}

@article{DeFelice:2010aj,
  author        = {De Felice, Antonio and Tsujikawa, Shinji},
  title         = {$f(R)$ Theories},
  journal       = {Living Rev. Relativ.},
  volume        = {13},
  number        = {1},
  pages         = {3},
  year          = {2010},
  doi           = {10.12942/lrr-2010-3},
  eprint        = {1002.4928},
  archivePrefix = {arXiv},
  primaryClass  = {gr-qc}
}

@article{RufSteinwachs2018a,
  author        = {Ruf, Michael S. and Steinwachs, Christian F.},
  title         = {One-Loop Divergences for $f(R)$ Gravity},
  journal       = {Phys. Rev. D},
  volume        = {97},
  number        = {4},
  pages         = {044049},
  year          = {2018},
  doi           = {10.1103/PhysRevD.97.044049},
  eprint        = {1711.04785},
  archivePrefix = {arXiv},
  primaryClass  = {gr-qc}
}

@article{Ruf:2017xon,
    author = "Ruf, Michael S. and Steinwachs, Christian F.",
    title = "{Quantum equivalence of $f(R)$ gravity and scalar-tensor theories}",
    eprint = "1711.07486",
    archivePrefix = "arXiv",
    primaryClass = "gr-qc",
    reportNumber = "FR-PHENO-2017-021",
    doi = "10.1103/PhysRevD.97.044050",
    journal = "Phys. Rev. D",
    volume = "97",
    number = "4",
    pages = "044050",
    year = "2018"
}

@article{Ohta:2017trn,
    author = "Ohta, Nobuyoshi",
    title = "{Quantum equivalence of $f(R)$ gravity and scalar{\textendash}tensor theories in the Jordan and Einstein frames}",
    eprint = "1712.05175",
    archivePrefix = "arXiv",
    primaryClass = "hep-th",
    reportNumber = "KU-TP-71, KU-TP 71",
    doi = "10.1093/ptep/pty008",
    journal = "PTEP",
    volume = "2018",
    number = "3",
    pages = "033B02",
    year = "2018"
}

@article{Dolgov:2003px,
    author = "Dolgov, A. D. and Kawasaki, Masahiro",
    title = "{Can modified gravity explain accelerated cosmic expansion?}",
    eprint = "astro-ph/0307285",
    archivePrefix = "arXiv",
    doi = "10.1016/j.physletb.2003.08.039",
    journal = "Phys. Lett. B",
    volume = "573",
    pages = "1--4",
    year = "2003"
}

@article{Faraoni:2006sy,
    author = "Faraoni, Valerio",
    title = "{Matter instability in modified gravity}",
    eprint = "astro-ph/0610734",
    archivePrefix = "arXiv",
    doi = "10.1103/PhysRevD.74.104017",
    journal = "Phys. Rev. D",
    volume = "74",
    pages = "104017",
    year = "2006"
}

@article{Deser:2007jk,
      author         = "Deser, Stanley and Woodard, R. P.",
      title          = "{Nonlocal Cosmology}",
      journal        = "Phys. Rev. Lett.",
      volume         = "99",
      year           = "2007",
      pages          = "111301",
      doi            = "10.1103/PhysRevLett.99.111301",
      eprint         = "0706.2151",
      archivePrefix  = "arXiv",
      primaryClass   = "astro-ph",
      reportNumber   = "UFIFT-QG-07-03, BRX-TH-589",
      SLACcitation   = "%%CITATION = ARXIV:0706.2151;%%"
}

@article{Deffayet:2009ca,
      author         = "Deffayet, C. and Woodard, R. P.",
      title          = "{Reconstructing the Distortion Function for Nonlocal
                        Cosmology}",
      journal        = "JCAP",
      volume         = "0908",
      year           = "2009",
      pages          = "023",
      doi            = "10.1088/1475-7516/2009/08/023",
      eprint         = "0904.0961",
      archivePrefix  = "arXiv",
      primaryClass   = "gr-qc",
      reportNumber   = "UFIFT-QG-08-07",
      SLACcitation   = "%%CITATION = ARXIV:0904.0961;%%"
}

@article{Nojiri:2010pw,
      author         = "Nojiri, Shin'ichi and Odintsov, Sergei D. and Sasaki,
                        Misao and Zhang, Ying-li",
      title          = "{Screening of cosmological constant in non-local
                        gravity}",
      journal        = "Phys. Lett.",
      volume         = "B696",
      year           = "2011",
      pages          = "278-282",
      doi            = "10.1016/j.physletb.2010.12.035",
      eprint         = "1010.5375",
      archivePrefix  = "arXiv",
      primaryClass   = "gr-qc",
      reportNumber   = "YITP-10-91",
      SLACcitation   = "%%CITATION = ARXIV:1010.5375;%%"
}

@article{Shapiro:2008sf,
      author         = "Shapiro, Ilya L.",
      title          = "{Effective Action of Vacuum: Semiclassical Approach}",
      journal        = "Class. Quant. Grav.",
      volume         = "25",
      year           = "2008",
      pages          = "103001",
      doi            = "10.1088/0264-9381/25/10/103001",
      eprint         = "0801.0216",
      archivePrefix  = "arXiv",
      primaryClass   = "gr-qc",
      SLACcitation   = "%%CITATION = ARXIV:0801.0216;%%"
}

@article{Giddings:2007pj,
      author         = "Giddings, Steven B.",
      title          = "{Black holes, information, and locality}",
      journal        = "Mod. Phys. Lett.",
      volume         = "A22",
      year           = "2007",
      pages          = "2949-2954",
      doi            = "10.1142/S0217732307025923",
      eprint         = "0705.2197",
      archivePrefix  = "arXiv",
      primaryClass   = "hep-th",
      SLACcitation   = "%%CITATION = ARXIV:0705.2197;%%"
}

@article{Wetterich:1997bz,
      author         = "Wetterich, Christof",
      title          = "{Effective nonlocal Euclidean gravity}",
      journal        = "Gen. Rel. Grav.",
      volume         = "30",
      year           = "1998",
      pages          = "159-172",
      doi            = "10.1023/A:1018837319976",
      eprint         = "gr-qc/9704052",
      archivePrefix  = "arXiv",
      primaryClass   = "gr-qc",
      reportNumber   = "HD-THEP-97-12",
      SLACcitation   = "%%CITATION = GR-QC/9704052;%%"
}

@article{Barnaby:2007ve,
      author         = "Barnaby, Neil and Kamran, Niky",
      title          = "{Dynamics with infinitely many derivatives: The Initial
                        value problem}",
      journal        = "JHEP",
      volume         = "02",
      year           = "2008",
      pages          = "008",
      doi            = "10.1088/1126-6708/2008/02/008",
      eprint         = "0709.3968",
      archivePrefix  = "arXiv",
      primaryClass   = "hep-th",
      SLACcitation   = "%%CITATION = ARXIV:0709.3968;%%"
}

@article{Koivisto:2008xf,
      author         = "Koivisto, Tomi and Mota, David F.",
      title          = "{Vector Field Models of Inflation and Dark Energy}",
      journal        = "JCAP",
      volume         = "0808",
      year           = "2008",
      pages          = "021",
      doi            = "10.1088/1475-7516/2008/08/021",
      eprint         = "0805.4229",
      archivePrefix  = "arXiv",
      primaryClass   = "astro-ph",
      SLACcitation   = "%%CITATION = ARXIV:0805.4229;%%"
}

@article{Koivisto:2008dh,
      author         = "Koivisto, Tomi S.",
      title          = "{Newtonian limit of nonlocal cosmology}",
      journal        = "Phys. Rev.",
      volume         = "D78",
      year           = "2008",
      pages          = "123505",
      doi            = "10.1103/PhysRevD.78.123505",
      eprint         = "0807.3778",
      archivePrefix  = "arXiv",
      primaryClass   = "gr-qc",
      SLACcitation   = "%%CITATION = ARXIV:0807.3778;%%"
}

@article{Deser:2013uya,
      author         = "Deser, S. and Woodard, R. P.",
      title          = "{Observational Viability and Stability of Nonlocal
                        Cosmology}",
      journal        = "JCAP",
      volume         = "1311",
      year           = "2013",
      pages          = "036",
      doi            = "10.1088/1475-7516/2013/11/036",
      eprint         = "1307.6639",
      archivePrefix  = "arXiv",
      primaryClass   = "astro-ph.CO",
      reportNumber   = "BRX-TH-6689, CALT-68-2946, UFIFT-QG-13-05",
      SLACcitation   = "%%CITATION = ARXIV:1307.6639;%%"
}

@article{delaCruzDombriz:2008cp,
      author         = "de la Cruz-Dombriz, A. and Dobado, A. and Maroto, Antonio
                        Lopez",
      title          = "{On the evolution of density perturbations in f(R)
                        theories of gravity}",
      journal        = "Phys. Rev.",
      volume         = "D77",
      year           = "2008",
      pages          = "123515",
      doi            = "10.1103/PhysRevD.77.123515",
      eprint         = "0802.2999",
      archivePrefix  = "arXiv",
      primaryClass   = "astro-ph",
      SLACcitation   = "%%CITATION = ARXIV:0802.2999;%%"
}

@article{Bamba:2008ut,
    author = "Bamba, Kazuharu and Nojiri, Shin'ichi and Odintsov, Sergei D.",
    title = "{The Universe future in modified gravity theories: Approaching the finite-time future singularity}",
    eprint = "0807.2575",
    archivePrefix = "arXiv",
    primaryClass = "hep-th",
    doi = "10.1088/1475-7516/2008/10/045",
    journal = "JCAP",
    volume = "10",
    pages = "045",
    year = "2008"
}

@article{Cotsakis:2023uyt,
    author = "Cotsakis, Spiros and Mimoso, Jose P. and Miritzis, John",
    title = "{The cosmological frame principle and cosmic acceleration}",
    eprint = "2304.12733",
    archivePrefix = "arXiv",
    primaryClass = "gr-qc",
    doi = "10.1140/epjc/s10052-023-11922-z",
    journal = "Eur. Phys. J. C",
    volume = "83",
    number = "8",
    pages = "735",
    year = "2023"
}

@article{HuSawicki2007,
  author        = {Hu, Wayne and Sawicki, Ignacy},
  title         = {Models of $f(R)$ Cosmic Acceleration that Evade Solar-System Tests},
  journal       = {Phys. Rev. D},
  volume        = {76},
  pages         = {064004},
  year          = {2007},
  doi           = {10.1103/PhysRevD.76.064004},
  eprint        = {0705.1158},
  archivePrefix = {arXiv},
  primaryClass  = {astro-ph}
}

@article{Song:2006ej,
  author        = {Song, Yong-Seon and Hu, Wayne and Sawicki, Ignacy},
  title         = {The Large Scale Structure of $f(R)$ Gravity},
  journal       = {Phys. Rev. D},
  volume        = {75},
  pages         = {044004},
  year          = {2007},
  doi           = {10.1103/PhysRevD.75.044004},
  eprint        = {astro-ph/0610532},
  archivePrefix = {arXiv}
}

@article{Nojiri:2006ri,
  author        = {Nojiri, Shin'ichi and Odintsov, Sergei D.},
  title         = {Modified $f(R)$ Gravity Consistent with Realistic Cosmology: From a Matter Dominated Epoch to a Dark Energy Universe},
  journal       = {Phys. Rev. D},
  volume        = {74},
  pages         = {086005},
  year          = {2006},
  doi           = {10.1103/PhysRevD.74.086005},
  eprint        = {hep-th/0608008},
  archivePrefix = {arXiv}
}

@article{Nojiri:2009kx,
  author        = {Nojiri, Shin'ichi and Odintsov, Sergei D. and Saez-Gomez, Diego},
  title         = {Cosmological Reconstruction of Realistic Modified $F(R)$ Gravities},
  journal       = {Phys. Lett. B},
  volume        = {681},
  pages         = {74--80},
  year          = {2009},
  doi           = {10.1016/j.physletb.2009.09.045},
  eprint        = {0908.1269},
  archivePrefix = {arXiv},
  primaryClass  = {hep-th}
}

@article{Nojiri:2010wj,
  author        = {Nojiri, Shin'ichi and Odintsov, Sergei D.},
  title         = {Unified Cosmic History in Modified Gravity: From $F(R)$ Theory to Lorentz Non-Invariant Models},
  journal       = {Phys. Rept.},
  volume        = {505},
  pages         = {59--144},
  year          = {2011},
  doi           = {10.1016/j.physrep.2011.04.001},
  eprint        = {1011.0544},
  archivePrefix = {arXiv},
  primaryClass  = {gr-qc}
}

@article{Olmo:2011uz,
    author = "Olmo, Gonzalo J.",
    title = "{Palatini Approach to Modified Gravity: f(R) Theories and Beyond}",
    eprint = "1101.3864",
    archivePrefix = "arXiv",
    primaryClass = "gr-qc",
    doi = "10.1142/S0218271811018925",
    journal = "Int. J. Mod. Phys. D",
    volume = "20",
    pages = "413--462",
    year = "2011"
}

@article{Harko:2011nh,
    author = "Harko, Tiberiu and Koivisto, Tomi S. and Lobo, Francisco S. N. and Olmo, Gonzalo J.",
    title = "{Metric-Palatini gravity unifying local constraints and late-time cosmic acceleration}",
    eprint = "1110.1049",
    archivePrefix = "arXiv",
    primaryClass = "gr-qc",
    doi = "10.1103/PhysRevD.85.084016",
    journal = "Phys. Rev. D",
    volume = "85",
    pages = "084016",
    year = "2012"
}

@article{Hehl:1994ue,
    author = "Hehl, Friedrich W. and McCrea, J. Dermott and Mielke, Eckehard W. and Ne'eman, Yuval",
    title = "{Metric affine gauge theory of gravity: Field equations, Noether identities, world spinors, and breaking of dilation invariance}",
    eprint = "gr-qc/9402012",
    archivePrefix = "arXiv",
    reportNumber = "TAUP-N192-94, TAUP-192-94",
    doi = "10.1016/0370-1573(94)00111-F",
    journal = "Phys. Rept.",
    volume = "258",
    pages = "1--171",
    year = "1995"
}

@article{Vilkovisky:1984st,
    author = "Vilkovisky, G. A.",
    title = "{The Unique Effective Action in Quantum Field Theory}",
    doi = "10.1016/0550-3213(84)90228-1",
    journal = "Nucl. Phys. B",
    volume = "234",
    pages = "125--137",
    year = "1984"
}

@book{DeWitt:2003pm,
  author    = {DeWitt, Bryce S.},
  title     = {The Global Approach to Quantum Field Theory},
  series    = {International Series of Monographs on Physics},
  volume    = {114},
  publisher = {Oxford University Press},
  address   = {Oxford},
  year      = {2003},
  isbn      = {978-0-19-851093-2}
}

@article{Kuntz:2026vhs,
    author = "Kuntz, Iber{\^e} and Liberati, Stefano",
    title = "{Off-shell equivalence in quantum field theory and gravity}",
    eprint = "2607.12644",
    archivePrefix = "arXiv",
    primaryClass = "hep-th",
    month = "7",
    journal="",
    year = "2026"
}

@article{Flanagan:2004bz,
    author = "Flanagan, Eanna E.",
    title = "{The Conformal frame freedom in theories of gravitation}",
    eprint = "gr-qc/0403063",
    archivePrefix = "arXiv",
    doi = "10.1088/0264-9381/21/15/N02",
    journal = "Class. Quant. Grav.",
    volume = "21",
    pages = "3817",
    year = "2004"
}

@article{Faraoni:2006fx,
    author = "Faraoni, Valerio and Nadeau, Shahn",
    title = "{The (pseudo)issue of the conformal frame revisited}",
    eprint = "gr-qc/0612075",
    archivePrefix = "arXiv",
    doi = "10.1103/PhysRevD.75.023501",
    journal = "Phys. Rev. D",
    volume = "75",
    pages = "023501",
    year = "2007"
}

@article{Kamenshchik:2014waa,
    author = "Kamenshchik, Alexander Yu. and Steinwachs, Christian F.",
    title = "{Question of quantum equivalence between Jordan frame and Einstein frame}",
    eprint = "1408.5769",
    archivePrefix = "arXiv",
    primaryClass = "gr-qc",
    reportNumber = "FR-PHENO-2014-011",
    doi = "10.1103/PhysRevD.91.084033",
    journal = "Phys. Rev. D",
    volume = "91",
    number = "8",
    pages = "084033",
    year = "2015"
}

@article{Bar:2013bpw,
    author = {B{\"a}r, Christian},
    title = "{Green-Hyperbolic Operators on Globally Hyperbolic Spacetimes}",
    eprint = "1310.0738",
    archivePrefix = "arXiv",
    primaryClass = "math-ph",
    doi = "10.1007/s00220-014-2097-7",
    journal = "Commun. Math. Phys.",
    volume = "333",
    number = "3",
    pages = "1585--1615",
    year = "2015"
}

@article{Salgado:2005hx,
    author = "Salgado, Marcelo",
    title = "{The Cauchy problem of scalar tensor theories of gravity}",
    eprint = "gr-qc/0509001",
    archivePrefix = "arXiv",
    doi = "10.1088/0264-9381/23/14/010",
    journal = "Class. Quant. Grav.",
    volume = "23",
    pages = "4719--4742",
    year = "2006"
}

@article{Dyer:2008hb,
    author = "Dyer, Ethan and Hinterbichler, Kurt",
    title = "{Boundary Terms, Variational Principles and Higher Derivative Modified Gravity}",
    eprint = "0809.4033",
    archivePrefix = "arXiv",
    primaryClass = "gr-qc",
    doi = "10.1103/PhysRevD.79.024028",
    journal = "Phys. Rev. D",
    volume = "79",
    pages = "024028",
    year = "2009"
}

@article{Guarnizo:2010xr,
    author = "Guarnizo, Alejandro and Castaneda, Leonardo and Tejeiro, Juan M.",
    title = "{Boundary Term in Metric f(R) Gravity: Field Equations in the Metric Formalism}",
    eprint = "1002.0617",
    archivePrefix = "arXiv",
    primaryClass = "gr-qc",
    doi = "10.1007/s10714-010-1012-6",
    journal = "Gen. Rel. Grav.",
    volume = "42",
    pages = "2713--2728",
    year = "2010"
}

@article{Capozziello:2007ms,
    author = "Capozziello, S. and Stabile, A. and Troisi, A.",
    title = "{The Newtonian Limit of f(R) gravity}",
    eprint = "0708.0723",
    archivePrefix = "arXiv",
    primaryClass = "gr-qc",
    doi = "10.1103/PhysRevD.76.104019",
    journal = "Phys. Rev. D",
    volume = "76",
    pages = "104019",
    year = "2007"
}

@article{Capozziello:2009vr,
    author = "Capozziello, S. and Stabile, A. and Troisi, A.",
    title = "{A General solution in the Newtonian limit of f(R)- gravity}",
    eprint = "0901.0448",
    archivePrefix = "arXiv",
    primaryClass = "gr-qc",
    doi = "10.1142/S0217732309030382",
    journal = "Mod. Phys. Lett. A",
    volume = "24",
    pages = "659--665",
    year = "2009"
}

@article{Stabile:2010zk,
    author = "Stabile, A.",
    title = "{The Post-Newtonian Limit of f(R)-gravity in the Harmonic Gauge}",
    eprint = "1004.1973",
    archivePrefix = "arXiv",
    primaryClass = "gr-qc",
    doi = "10.1103/PhysRevD.82.064021",
    journal = "Phys. Rev. D",
    volume = "82",
    pages = "064021",
    year = "2010"
}

@article{Berry:2011pb,
    author = "Berry, Christopher P. L. and Gair, Jonathan R.",
    title = "{Linearized f(R) Gravity: Gravitational Radiation and Solar System Tests}",
    eprint = "1104.0819",
    archivePrefix = "arXiv",
    primaryClass = "gr-qc",
    doi = "10.1103/PhysRevD.83.104022",
    journal = "Phys. Rev. D",
    volume = "83",
    pages = "104022",
    year = "2011",
    note = "[Erratum: Phys.Rev.D 85, 089906 (2012)]"
}

@article{Khoury:2003aq,
    author = "Khoury, Justin and Weltman, Amanda",
    title = "{Chameleon fields: Awaiting surprises for tests of gravity in space}",
    eprint = "astro-ph/0309300",
    archivePrefix = "arXiv",
    doi = "10.1103/PhysRevLett.93.171104",
    journal = "Phys. Rev. Lett.",
    volume = "93",
    pages = "171104",
    year = "2004"
}

@article{Khoury:2003rn,
    author = "Khoury, Justin and Weltman, Amanda",
    title = "{Chameleon cosmology}",
    eprint = "astro-ph/0309411",
    archivePrefix = "arXiv",
    doi = "10.1103/PhysRevD.69.044026",
    journal = "Phys. Rev. D",
    volume = "69",
    pages = "044026",
    year = "2004"
}

@article{Faraoni:2007yn,
    author = "Faraoni, Valerio",
    title = "{de Sitter space and the equivalence between f(R) and scalar-tensor gravity}",
    eprint = "gr-qc/0703044",
    archivePrefix = "arXiv",
    doi = "10.1103/PhysRevD.75.067302",
    journal = "Phys. Rev. D",
    volume = "75",
    pages = "067302",
    year = "2007"
}

@article{Barvinsky:2014lja,
    author = "Barvinsky, A. O.",
    title = "{Aspects of Nonlocality in Quantum Field Theory, Quantum Gravity and Cosmology}",
    eprint = "1408.6112",
    archivePrefix = "arXiv",
    primaryClass = "hep-th",
    doi = "10.1142/S0217732315400039",
    journal = "Mod. Phys. Lett. A",
    volume = "30",
    number = "03n04",
    pages = "1540003",
    year = "2015"
}

@article{Simon:1990ic,
    author = "Simon, Jonathan Z.",
    title = "{Higher Derivative Lagrangians, Nonlocality, Problems and Solutions}",
    reportNumber = "UCSB-TH-89-50",
    doi = "10.1103/PhysRevD.41.3720",
    journal = "Phys. Rev. D",
    volume = "41",
    pages = "3720",
    year = "1990"
}

@article{Burgess:2003jk,
    author = "Burgess, C. P.",
    title = "{Quantum gravity in everyday life: General relativity as an effective field theory}",
    eprint = "gr-qc/0311082",
    archivePrefix = "arXiv",
    doi = "10.12942/lrr-2004-5",
    journal = "Living Rev. Rel.",
    volume = "7",
    pages = "5--56",
    year = "2004"
}

@article{Solomon:2017nlh,
    author = "Solomon, Adam R. and Trodden, Mark",
    title = "{Higher-derivative operators and effective field theory for general scalar-tensor theories}",
    eprint = "1709.09695",
    archivePrefix = "arXiv",
    primaryClass = "hep-th",
    doi = "10.1088/1475-7516/2018/02/031",
    journal = "JCAP",
    volume = "02",
    pages = "031",
    year = "2018"
}

@article{Hindawi:1995an,
    author = "Hindawi, Ahmed and Ovrut, Burt A. and Waldram, Daniel",
    title = "{Consistent spin two coupling and quadratic gravitation}",
    eprint = "hep-th/9509142",
    archivePrefix = "arXiv",
    reportNumber = "UPR-0660-T, UPR-660T",
    doi = "10.1103/PhysRevD.53.5583",
    journal = "Phys. Rev. D",
    volume = "53",
    pages = "5583--5596",
    year = "1996"
}

@article{Rodrigues:2011zi,
    author = "Rodrigues, Davi C. and de O. Salles, Filipe and Shapiro, Ilya L. and Starobinsky, Alexei A.",
    title = "{Auxiliary fields representation for modified gravity models}",
    eprint = "1101.5028",
    archivePrefix = "arXiv",
    primaryClass = "gr-qc",
    reportNumber = "RESCEU-25-10",
    doi = "10.1103/PhysRevD.83.084028",
    journal = "Phys. Rev. D",
    volume = "83",
    pages = "084028",
    year = "2011"
}

@article{Babichev:2019twf,
    author = "Babichev, Eugeny and Izumi, Keisuke and Tanahashi, Norihiro and Yamaguchi, Masahide",
    title = "{Invertible field transformations with derivatives: necessary and sufficient conditions}",
    eprint = "1907.12333",
    archivePrefix = "arXiv",
    primaryClass = "hep-th",
    reportNumber = "LPT-Orsay-19-39",
    doi = "10.4310/ATMP.2021.v25.n2.a2",
    journal = "Adv. Theor. Math. Phys.",
    volume = "25",
    number = "2",
    pages = "309--325",
    year = "2021"
}

@article{Babichev:2021bim,
    author = "Babichev, Eugeny and Izumi, Keisuke and Tanahashi, Norihiro and Yamaguchi, Masahide",
    title = "{Invertibility conditions for field transformations with derivatives: Toward extensions of disformal transformation with higher derivatives}",
    eprint = "2109.00912",
    archivePrefix = "arXiv",
    primaryClass = "hep-th",
    doi = "10.1093/ptep/ptab151",
    journal = "PTEP",
    volume = "2022",
    number = "1",
    pages = "013A01",
    year = "2022"
}

\end{document}